\documentclass{aa}  
\usepackage{amssymb}
\usepackage{graphicx}
\usepackage{color}
\usepackage{rotating}
\usepackage{pdflscape}
\usepackage{lscape}
\usepackage{float}
\usepackage{txfonts}
\newcommand{\teff}{\ifmmode T_{\rm eff} \else T$_{\mathrm{eff}}$~\fi}
\newcommand{\logg}{\ifmmode \log g \else $\log g$~\fi}
\newcommand{\lL}{\ifmmode \log(L/L_{\odot}) \else $\log(L/L_{\odot})$~\fi}

\newcommand{\vsini}{$\upsilon$ sin$i$}

\newcommand{\vturb}{$\xi_{\rm turb}$}

\newcommand{\kms}{km s$^{-1}$~}
\newcommand{\msun}{\ifmmode M_{\odot} \else M$_{\odot}$~\fi}
\newcommand{\zsun}{\ifmmode Z_{\odot} \else Z$_{\odot}$~\fi}
\newcommand{\lsun}{\ifmmode L_{\odot} \else L$_{\odot}$~\fi}
\newcommand{\rsun}{\ifmmode R_{\odot} \else R$_{\odot}$~\fi}
\newcommand{\qh}{\ifmmode Q_{\rm H} \else $Q_{\rm H}$~\fi}
\newcommand{\ciso}{\ifmmode ^{12}{\rm C}/^{13}{\rm C} \else $^{12}{\rm C}/^{13}{\rm C}$~\fi}
\newcommand{\qhei}{\ifmmode Q_{\ion{He}{i}} \else $Q_{\ion{He}{i}}$\fi}

\newcommand{\mum}{\ifmmode \mu m \else $\mu m$\fi}

\begin{document}

\title{New determination of abundances and stellar parameters for a set of weak G-band stars
\thanks{Based on data collected at La Silla Observatory (ESO, Chile), program identifier ID 089.D-0189(A),
and at the Telescope Bernard Lyot (USR5026) operated by the Observatoire Midi-Pyrénées, Université de Toulouse (Paul Sabatier), Centre National de la Recherche Scientifique of France.}
}
   \author{
          A. Palacios\inst{1}
          \and
          G. Jasniewicz\inst{1} 
          \and
          T. Masseron\inst{2}
          \and
          F. Th\'evenin\inst{3}
          \and
          J. Itam-Pasquet\inst{1}
          \and
          M. Parthasarathy\inst{4}
          }

   \offprints{A. Palacios: ana.palacios AT umontpellier.fr}

   \institute{LUPM UMR 5299 CNRS/UM, Universit\'e de Montpellier, CC 72, F-34095 Montpellier Cedex 05, France
         \and
        Institute of Astronomy - University of Cambridge, Madingley Road, Cambridge CB3 0HA,  United Kingdom
         \and
             Laboratoire Lagrange, UMR 7293, OCA, CS 34229, F-06304 Nice Cedex 4, France 
         \and    
              Indian Institute of Astrophysics, Koramangala, Bangalore - 560034, India
             }

   \date{}

\authorrunning{A. Palacios et al.}
\titlerunning{Abundances and stellar parameters for a set of wGb stars}

\date{Received, accepted}

\abstract
   {Weak G-band (wGb) stars are a very peculiar class of red giants; they are almost devoided of carbon and often present mild lithium enrichment. Despite their very puzzling abundance patterns, very few detailed spectroscopic studies existed up to a few years ago, which prevented any clear understanding of the wGb phenomenon. We recently proposed the first consistent analysis of published data for a sample of 28 wGb stars and were able to identify them as descendants of early A-type to late B-type stars, although we were not able to conclude on their evolutionary status or the origin of their peculiar abundance pattern.}
  {Using new high-resolution spectra, we present the study of a new sample of wGb stars with the aim of homogeneously deriving their fundamental parameters and surface abundances for a selected set of chemical species that we use to improve our insight on this peculiar class of objects.}
  {We  obtained high-resolution and high signal-to-noise spectra for 19 wGb stars in the southern and northern hemisphere that we used to perform consistent spectral synthesis to derive their fundamental parameters and  metallicities, as well as the spectroscopic abundances for Li, C, $^{12}$C/$^{13}$C, N, O, Na, Sr, and Ba. We also computed dedicated stellar evolution models that we used to determine the masses and to investigate the evolutionary status and chemical history of the stars in our sample. }
  {We confirm that the wGb stars are stars with initial masses in the range 3.2 to 4.2 \msun. We suggest that a large fraction  could be mildly evolved stars on the subgiant branch currently undergoing the first dredge-up, while a smaller number of stars are more probably in the core He burning phase at the clump. After analysing their abundance pattern, we confirm their strong nitrogen enrichment anti-correlated with large carbon depletion, characteristic of material fully processed through the CNO cycle to an extent not known in evolved intermediate-mass stars in the field and in open clusters. However, we demonstrate here that such a pattern is very unlikely owing to self-enrichment.}
  {In the light of the current observational constraints, no solid self-consistent pollution scenario can be presented either, leaving the wGb puzzle largely unsolved. }


\maketitle

\section{Introduction}\label{sec:intro}

The weak G-band (wGb) stars are G and K giants whose spectra show very weak or
absent G-band of the molecule CH at 4300\AA. The first one, HD18474, was discovered in the northern sky by \citet{bidelman1951}
and is the prototype of this class of stars. Later \citet{bidelman1973} established a list of 34 wGb stars in the southern
sky (declination -83$^\circ$ < $\delta < +6^\circ$). These stars are rare among the population of G-K giants in the Galaxy.
The ratio of wGb stars among G-K giants in the Bright Star Catalogue is less than 0.3\%. 
This small sample of stars was mainly studied in the late 1970s and early 1980s \citep[see e.g.][]{sneden1978,rao1978,cottrell1978,partha1980,day1980,partha1984}. Later \citet{sneden1984} found an excellent agreement between carbon abundances derived from spectroscopic CO data and those determined using features of the CH (G-band). \citet{lambert1984} confirmed the presence of lithium in some of the wGb stars, adding yet another complexity to these peculiar objects. It is only in the past few years that the interest for these objects has been reawakened with the papers by \citet{palaciosWGB2012} (hereafter {\sf Paper I}) and
\citet{AL13} (hereafter {\sf AL13}).

\begin{table*}[t]
\caption{Basic data and atmospheric parameters of weak G-band stars. For rotational velocities \vsini, 
the typical error is 2km.s$^{-1}$} 
\centering
\begin{tabular}{rcccccccccc}
\hline
HD No   &  T   & eT  &logg & elogg & \lL & e\lL & \vturb  & e\vturb &\vsini & colour\\         
& (K) & (K) & (dex) & (dex) & (dex) & (dex)  & km.s$^{-1}$ & km.s$^{-1}$  & km.s$^{-1}$ & coding \\         
\hline
\hline
HD 18474  &5198  & 38 &2.65&0.03& 2.11&0.08&1.54&0.03  & $<5$ &  black\\  
HD 49960  &5030  & 32 &2.61&0.05& 2.09&0.09&1.89&0.05  & $<5$ &  red  \\  
HD 56438  &5037  &163 &2.75&0.01& 1.94&0.09&1.15&0.11  & $<5$ &  green  \\  
HD 67728  &4827  &121 &2.28&0.13& 2.42&0.16&1.87&0.07  & 13 &  blue  \\  
HD 78146  &4734  & 96 &2.13&0.03& 2.55&0.09&1.74&0.05  & $<5$ &  cyan  \\  
HD 82595  &4995  & 79 &2.28&0.01& 2.47&0.08&1.62&0.04  & $<5$ &  magenta \\   
HD 91805  &5247  & 15 &2.56&0.04& 2.21&0.08&1.63&0.03  & $<5$ &  orange  \\  
HD 94956  &5131  & 75 &2.76&0.30& 1.99&0.31&1.74&0.04  & $<5$ &  light blue  \\  
HD102851  &4991  & 41 &2.68&0.26& 2.01&0.27&1.91&0.05  & $<5$ &  reddish brown  \\  
HD119256  &4984  & 54 &2.64&0.07& 2.04&0.10&1.71&0.04  & $<5$ &   light pink\\  
HD120170  &5127  & 38 &2.76&0.06& 1.97&0.09&1.46&0.03  & $<5$ &  light brown \\  
HD120213  &4577  & 46 &1.95&0.26& 2.71&0.27&1.70&0.05  & $<5$ &  light green \\  
HD124721  &5107  & 61 &2.64&0.11& 2.10&0.13&1.58&0.04  & $<5$ &  olive green  \\  
HD146116  &4920  & 33 &2.03&0.41& 2.73&0.42&2.04&0.05  & $<5$ &  violet  \\  
HD165462  &5078  &110 &2.45&0.13& 2.30&0.15&1.72&0.04  & 11 &  dark red  \\  
HD165634  &5114  & 45 &2.55&0.08& 2.19&0.11&1.54&0.03  & $<5$ &  lime green  \\  
HD166208  &5177  & 52 &2.81&0.04& 1.94&0.08&1.92&0.05  & $<5$ &  golden yellow  \\  
HD204046  &4984  & 90 &2.51&0.37& 2.20&0.38&1.76&0.04  & $<5$ &  purple  \\  
HD207774  &5125  & 33 &2.79&0.42& 1.96&0.43&1.85&0.04  & $<5$ &  grey  \\  
 \hline
\end{tabular} 
\label{tab_AP}        
\end{table*}                                                

In {\sf paper I}, we homogeneously reanalysed spectroscopic archival data for a sample of wGb stars, and used dedicated stellar evolution models to establish their mass range. They appear to be the descendants of 3 M$_\odot$ to 4.5 M$_\odot$ stars, now located in the crowded Hertzsprung-Russell diagram region of the red clump, which makes their evolutionary status ambiguous (red giant or core helium burning stars). This result is confirmed by {\sf AL13} in their new spectroscopic study of a large sample of 24 southern wGb stars.\\ They establish that wGb stars present a strong depletion in carbon that is about a factor of 20 larger than for normal giants and that the $^{12}$C/$^{13}$C ratio approaches the CN-cycle equilibrium value. In addition, they emphasize a strong N overabundance anti-correlated with the C underabundance, which indicates that the atmospheres of wGb stars have been processed through the CN-cycle and probably also the ON-cycle. These abundance anomalies are often accompanied by high lithium abundances, similar to those of Li-rich K giants (A(Li) $\geq$ 1.4 dex).\\
The concurrence of high lithium content and CN(O) cycled material is extremely puzzling since these nuclei are issued from exclusive regions. The interpretation of the abundance pattern of wGb stars differs in {\sf AL13} and {\sf Paper I}, and both studies are inconclusive. 

 The goal of this paper is to perform a new homogeneous abundance analysis of a large sample of wGb stars with up-to-date model atmospheres and line-lists, and to compare them with the predictions of dedicated stellar evolution models in order to clarify the origin of the peculiarities of wGb stars.
To this end  we have conducted new spectroscopic observations of a sample of 19 wGb stars, mainly in the southern hemisphere.\\

 We present the sample and the observations in \S~\ref{sec:obs}, and the spectral analysis applied to retrieve the atmospheric parameters and abundances in \S~\ref{sec:spec};  we also compare our results to {\sf AL13} for the stars in common between the two samples. We discuss the dynamical properties of the wGb stars in \S~\ref{sec:kinematics}. We finally use dedicated up-to-date stellar evolution models to reinvestigate the masses and evolutionary status of the stars in our sample (\S~\ref{sec:evol}) and proceed to a discussion of the possible scenarios to account for their abundance pattern in \S~\ref{sec:scenario} before concluding in \S~\ref{sec:conclusion}.

\section{Observations}\label{sec:obs}
This paper is based on new observations of 19 wGb stars. 
Seventeen southern wGb stars were observed at La Silla, ESO Chile, with the high-efficiency Fiber-fed Extended Range Optical Spectrograph FEROS spectrograph mounted on the 2.2m telescope.
FEROS is a bench-mounted, thermally controlled, prism-cross-dispersed \'echelle spectrograph, providing, in a single spectrogram spread over 
39 orders, almost complete spectral coverage from $\sim350$ to $\sim920$~nm at a resolution of 48\,000 \citep{FEROS}. The FEROS observations were carried out 
during an observing run between May 10 and 13, 2012. 
All these spectra were flat-fielded and calibrated by means of ThArNe exposures using standard processing tools available
at ESO.\\
In addition, two northern wGb stars, HD\,18474 and HD\,166208, were observed in service mode at the Observatoire du Pic du Midi, France,
with the NARVAL spectrograph mounted on the Bernard Lyot 2.0m telescope. The NARVAL instrument consists
of a bench-mounted cross-dispersed
\'echelle spectrograph,
fibre-fed from a Cassegrain-mounted polarimeter unit \citep{NARVAL}. It was used in its non-polarimetric mode; it provided almost complete spectral coverage from $\sim375$ to $\sim1050$nm at a
resolution of 75\,000 in a single spectrogram spread over 40 orders, . The  NARVAL data were reduced using the data reduction software Libre-ESpRIT, written and provided by \citet{donati97} from the LATT (Observatoire Midi-Pyrénées).\\
 The observation log is given in Appendix (Table~\ref{tab_log_observations}).\\

\section{Spectral analysis}\label{sec:spec}

All spectra were processed with the   Brussels Automatic Code for Characterizing High
accUracy Spectra (BACCHUS) pipeline developed by T. Masseron. 
The BACCHUS code consists of three different modules  designed to derive equivalent widths (EWs), 
stellar parameters, and abundances. The current version relies on a grid of MARCS model atmospheres \citep{Gustafsson2008}, 
a specific procedure for interpolating the model atmosphere thermodynamical structure within the grid \citep{Masseron2006} 
and the radiative transfer code TURBOSPECTRUM \citep{Alvarez1998,Plez2012}.  It has been successfully compared to other spectral synthesis codes as shown in \citet{Jofre14} within the framework of the Gaia preparation team.

%% Figure 1
\begin{figure*}[ht]
\begin{center}
\includegraphics[height=0.8\textwidth,angle=-90]{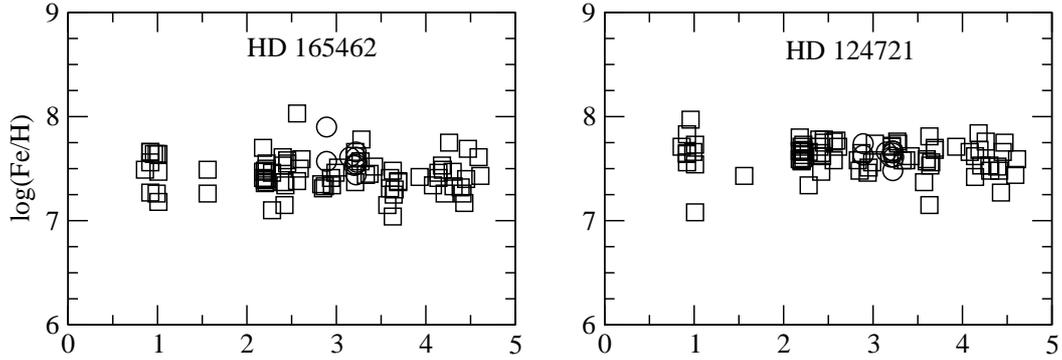}
\caption{Variation of the derived iron abundances as a function of the  excitation energy for two wGb stars of our program. Squares indicate Fe I lines and circles Fe II lines. 
}
\label{fig:FeAbund_khi}
\end{center}
\end{figure*}
%% Figure 
 
One asset of the BACCHUS code is the computation of EWs, because the
EW of only the considered line is taken into account
(excluding the contribution from blending lines).
Indeed, EWs are not computed  directly on the observed spectrum, but
rather from the synthetic spectrum with the best-matching abundance.
This way, we have access to the information about the contribution of
blending lines, which allows a clean measurement of the EW of the line of
interest.
As a consequence of this efficient blend-removal technique, BACCHUS 
tends to have EW measurements that are sometimes
slightly lower than  average, especially below 5000~\AA ~where blends are
numerous. 
The process measuring EWs is iterated with a new model
atmosphere taking into account the possible change in metallicity and
adjusting the microturbulence velocity by forcing the invariance of
derived abundances with respect to the EWs.
Additionally, a new convolution parameter for the spectral synthesis
encompassing macroturbulence velocity, instrument resolution, and
stellar rotation is determined and possibly adopted. This procedure
usually converges after two iterations when the
correction on the metallicity becomes smaller than the line abundance
dispersion.

The last step of the procedure consists in injecting $\sim$ 80 selected Fe lines and their derived 
 equivalent widths in TURBOSPECTRUM to derive abundances for a set of model atmospheres. 
For each model, the slopes of abundances against excitation potential and against equivalent widths, 
as well as Fe I and Fe II lines abundances,   are computed.   
The final parameters are determined by requesting that the ionization   equilibrium is fulfilled, 
and that simultaneously null slopes for abundances against excitation potential and against 
equivalent widths are encountered.

\subsection{Atmospheric parameters}\label{sec:APs}

Atmospheric parameters of our list of wGb stars are given in Table~\ref{tab_AP}.
Using the {\sf param} module of BACCHUS, the effective
temperature \teff, the logarithmic surface gravity \logg, the global
metallicity [Fe/H], and the microturbulent velocity \vturb~ are computed by removing
any trends in the [Fe/H] vs. $E$ and [Fe/H] vs. $W/L$ relations (where
$E$ is the energy of the lower level of a line, $W$ the measured equivalent
width, and $L$ the wavelength of the considered line), and by
forcing lines of Fe I and Fe II to yield the same abundance. 
The slopes of Fe  abundances against excitation potential
are given in Fig.~\ref{fig:FeAbund_khi} for two wGb stars. 
The abundance of an element like Fe I, as derived from different spectral lines in a wide range in
excitation potential, should be independent of the excitation potential if \teff is well determined. FeI and FeII
also must have similar abundances if the surface gravity \logg is well determined. This check for two stars
confirms that BACCHUS has robust results. There is some spread in Fig.~\ref{fig:FeAbund_khi}; the most marginal points in this plot disappear as soon as
$W/L <0.06$.

The projected rotational velocity  \vsini~ is determined for each star by matching a rotational profile to the average line broadening. The luminosities in Table~\ref{tab_AP} are derived from basic expressions of luminosity and gravity using the relation
\begin{equation}
 \log(L/L_{\odot})  =  -10.61 +  \log(M/M_{\odot}) - \log(g)  + 4\log(T_\mathrm{eff})
\label{logL}
,\end{equation}

where $M$ is the mass of the star in solar units.   
For the calculation of $\log(L/L_{\odot}) $ we use the masses derived from our stellar evolution models (see \S~\ref{sec:evol}) and reported in Table~\ref{tab_MRage}.

%% Figure 2
\begin{figure}
\begin{center}
\includegraphics[width=0.4\textwidth,angle=-90]{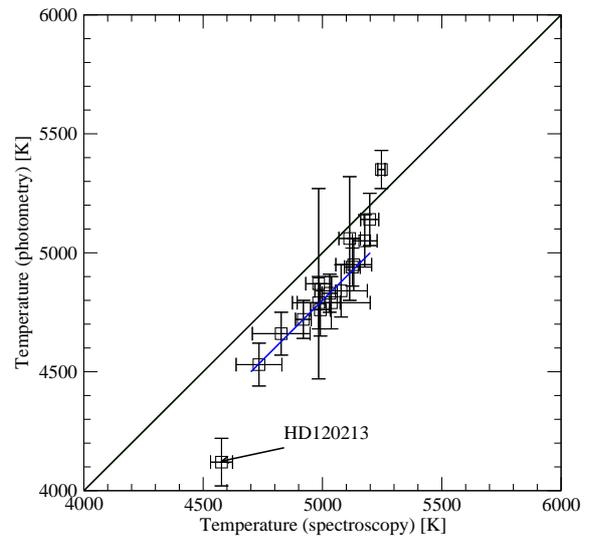}
\caption{Comparison of spectroscopic and photometric temperatures. The spectroscopic temperatures are consistently higher than the photometric values. The small light segment indicates an offset of 200 K from the 1:1 ratio shown by the black line.}
\label{fig:TphotTspec}
\end{center}
\end{figure}
%% Figure 1

Fig.~\ref{fig:TphotTspec} shows a clear discrepancy between our spectroscopic and photometric temperatures; this discrepancy is comparable to that found by AL13 for their own data
(see their Fig.~1 lower panel).
 The photometric temperatures (see \citet{palaciosWGB2012}, Table 1 for stars in common with that paper 
and the present study) are systematically lower by about 150 $K$ (standard deviation 133 $K$) compared
 to the spectroscopic values (given in Table~\ref{tab_AP}). This point will be discussed again in \S~\ref{sec:CNpec}.
 We note a significant difference (about 400 $K$) for HD120213, which is the coldest wGb star
in our sample.
In addition, there is a general good agreement within the uncertainties (0.2dex), except for HD91805 (0.5dex),
 between the luminosities determined  from photometry and parallaxes
in \citet{palaciosWGB2012} and those determined by spectroscopy in this work (see Table~\ref{tab_AP}).

%% Figure 3
\begin{figure}[H]
\begin{center}
\includegraphics[width=0.45\textwidth]{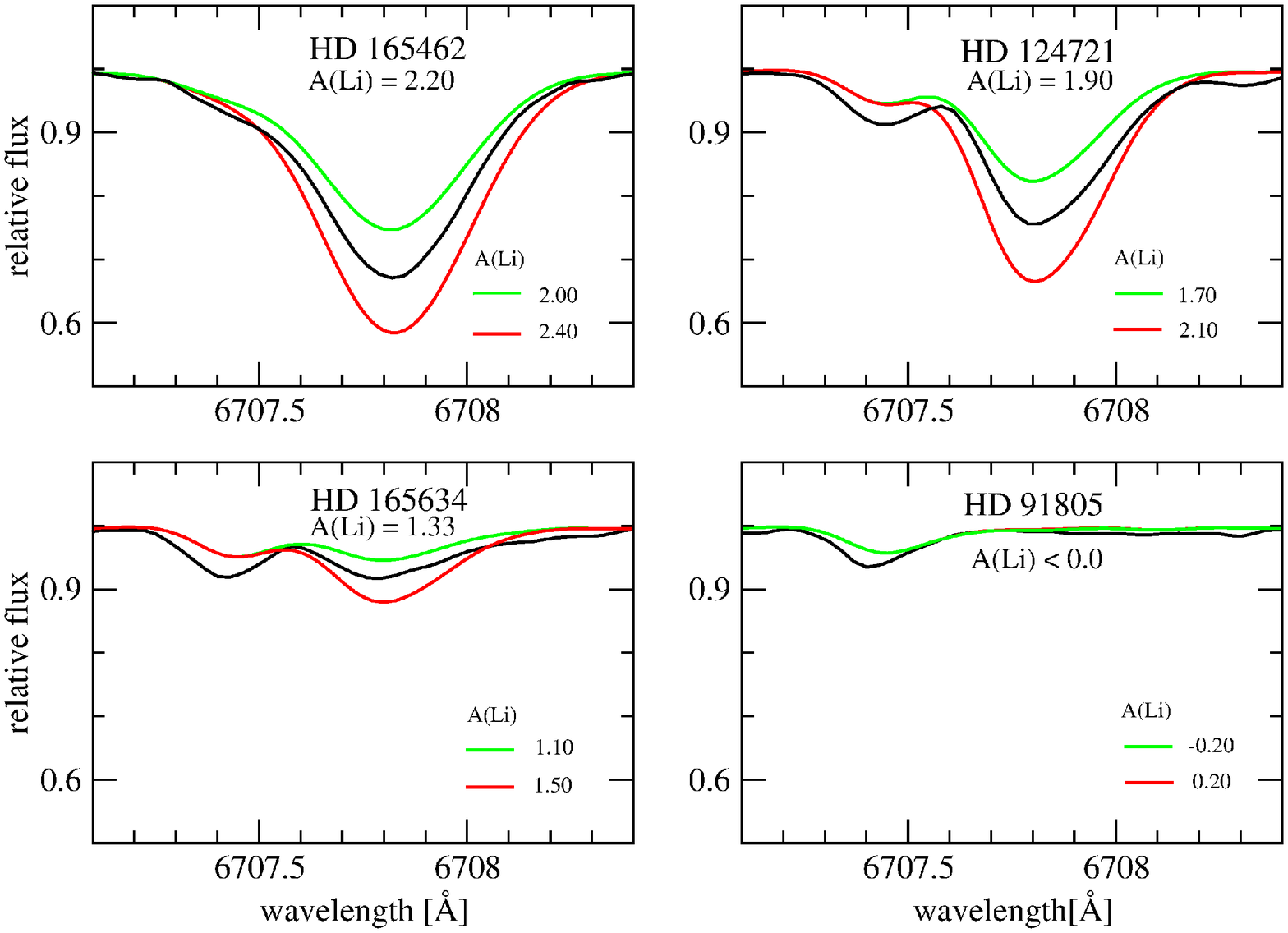}
%\hspace{1cm}
\includegraphics[width=0.45\textwidth]{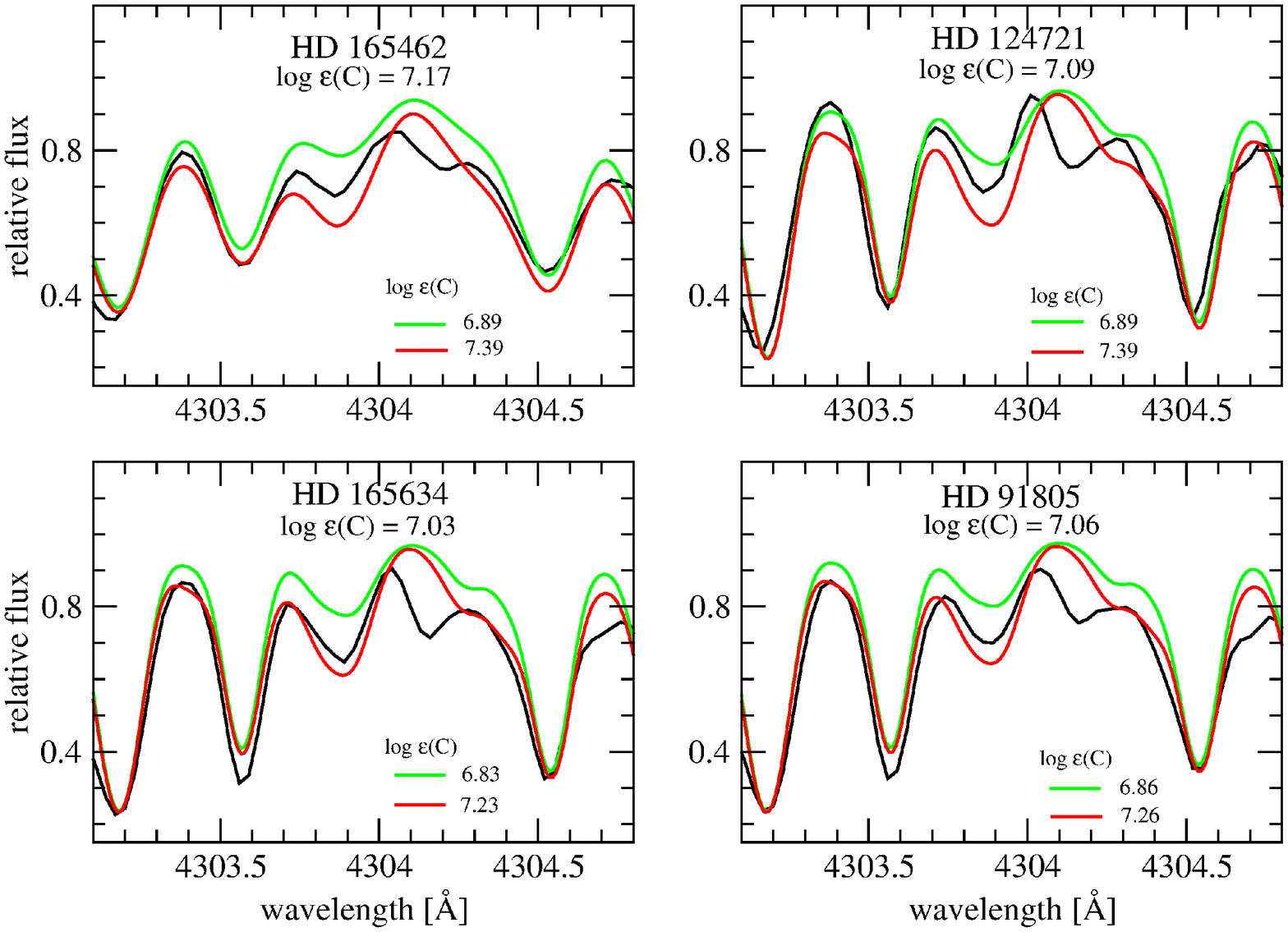}
\caption{Spectra of four wGb stars of our program stars and overplotted synthetic spectra around the 6707.8\AA~ Li I resonance doublet ({\em upper panel}) and the 4304\AA~ C line within the CH G band ({\em left panel}) . }
\label{fig:abund_LiC}
\end{center}
\end{figure}
%% Figure 3

\begin{table*}[ht]
\caption{Abundances  of light elements in wGb stars. All element abundances are given with respect to H, except for lithium
which is given in $\log \epsilon(Li)$. Solar values are taken from \citet{AGS05} and are 8.39, 7.78, 8.66, and 6.17 for C, N, O, and Na, respectively.
}  \label{tab_abund1}
\centering
\begin{tabular}{lcccccccccccc}
\hline
HD No   & [C/H] & e[C/H] & [N/H]& e[N/H]  & [O/H] & e[O/H] & [Na/H] & e[Na/H] & A(Li) & eLi & $^{12}$C/$^{13}$C & e$^{12}$C/$^{13}$C  \\         
\hline
\hline
HD 18474 &-1.65 &0.14&1.01 & 0.07&  0.09 &0.05&+0.38  &0.03& 1.58     &0.05& :     & :  \\  
HD 49960 &-1.42 &0.11&1.13 & 0.09&  0.24 &0.09&+0.25  &0.04& 0.90     &0.05&10     & 3     \\  
HD 56438 &-1.58 &0.11&1.06 & 0.08& -0.10 &0.05& 0.30  &0.07& 2.03     &0.05& 5     & 2     \\  
HD 67728 &-1.22 &0.09&1.24 & 0.06&  0.06 &0.05&-0.20  &0.12& $\leq0.0$&0.20& 4     & 1     \\  
HD 78146 &-1.40 &0.10&1.16 & 0.07& -0.06 &0.06&+0.25  &0.05& 1.01     &0.06& 4     & 1     \\  
HD 82595 &-1.38 &0.09&1.09 & 0.07& -0.12 &0.05&+0.31  &0.09& 1.43     &0.05& 5     & 1    \\   
HD 91805 &-1.33 &0.11&1.19 & 0.09&  0.11 &0.05&+0.37  &0.06& $\leq0.0$&0.20& 5     & 2     \\  
HD 94956 &-1.31 &0.09&1.13 & 0.06&  0.10 &0.05&+0.30  &0.05& 1.19     &0.06& 4     & 1     \\  
HD 102851&-1.52 &0.11&1.36 & 0.06&  0.52 &0.05&+0.10  &0.08& 0.50     &0.20& 4     & 1     \\  
HD 119256&-1.22 &0.10&1.20 & 0.05&  0.22 &0.06&+0.23  &0.10& 1.07     &0.08& 4     & 1     \\  
HD 120170&-1.60 &0.11&1.10 & 0.13&  0.06 &0.05&+0.01  &0.08& 2.92     &0.05& 5     & 2     \\  
HD 120213&-1.54 &0.10&0.95 & 0.06&  0.29 &0.11&+0.16  &0.07& 1.97     &0.05& 5     & 2     \\  
HD 124721&-1.30 &0.07&1.13 & 0.07&  0.23 &0.09& 0.30  &0.07& 1.90     &0.09& 5     & 2     \\  
HD 146116&-1.35 &0.10&0.93 & 0.06& -0.17 &0.05&+0.09  &0.05& $\leq0.0$&0.20& 5     & 1     \\  
HD 165462&-1.22 &0.06&1.18 & 0.06&  0.00 &0.05& 0.39  &0.02& 2.20     &0.05& 5     & 2     \\  
HD 165634&-1.36 &0.10&1.09 & 0.06&  0.12 &0.05& 0.39  &0.07& 1.33     &0.06& 5     & 2     \\  
HD 166208&-0.94 &0.09&1.70 & 0.06& -0.04 &0.08& 0.46  &0.07& 1.90     &0.06& 8     & 2     \\  
HD 204046&-1.40 &0.10&1.10 & 0.07&  0.08 &0.06& 0.03  &0.09& 0.67     &0.07& 5     & 1     \\  
HD 207774&-1.20 &0.12&1.06 & 0.06&  0.10 &0.06& 0.13  &0.05& $\leq0.0$&0.20& 4     & 1     \\  
 \hline 
\end{tabular} 
\end{table*}

\subsection{Abundance determinations}

Abundances were determined by means of the {\sf abund} module of
BACCHUS. The abundance determination module proceeds in the following way: 
(i) a spectrum synthesis, using the full set of lines  (atomic and molecular), is used for local continuum level finding (correcting for a
possible spectrum slope);
(ii) cosmic and telluric rejections are performed;
(iii) local signal-to-noise is estimated; and 
(iv) a series of flux points contributing to a given
absorption line is selected.
Abundances are then derived by comparison of the observed spectrum
with a set of convolved synthetic spectra characterized by different
abundances.  
Four different diagnostics are used: $\chi^2$ fitting, core line intensity
comparison, global goodness-of-fit estimate, and equivalent width
comparison. 
A decision tree then rejects the line or accepts it keeping the best-matching abundance.

  After the software automatically accepts a line, we then select the lines for which three diagnostics out of four (see
 above) are fulfilled, the order of priority being $\chi^2$ fitting,
 core line intensity comparison, and equivalent width comparison. We then
 perform a visual inspection of each line in the spectrum
 together with a set of overplotted synthetic spectra with different
 abundances.
 
 Examples of fits to lithium and carbon lines are given in Fig.~\ref{fig:abund_LiC}.
 
  The final derived abundances are given in
 Table~\ref{tab_abund1} and Table~\ref{tab_abund2}. Solar abundances
 are taken from \citet{AGS05}. The  most
 relevant lines for each element that have been used for the
 determination of LTE abundances are listed here:
\begin{itemize}
\item Lithium : 6707.8\AA~ (see Fig.~\ref{fig:abund_LiC})
\item Nitrogen : about 30 lines, from 7874.9\AA~  to 8060.2\AA 
\item Carbon : two lines in the CH G band at 4304\AA~(see Fig~\ref{fig:abund_LiC}) and 4311\AA, and two atomic lines in the red part at 8058.6\AA~ and 8335.1\AA 
\item Oxygen : 6300.3\AA 
\item Sodium : 4664.8\AA, 5148.8\AA, 5682.6\AA 
\item Strontium : 4161.8\AA, 4962.3\AA, 7070.1\AA 
\item Barium : 4166\AA, 4554\AA, 5853.7\AA 
\end{itemize}
 
We note that these stars are not metal-poor and are poorly affected by NLTE effects on both the determination of \logg by ionisation balance and on the chemical abundance determination \citep{TI99,IT00}. \\

We confirm the strong carbon deficiency in the wGb stars observed and a carbon isotopic ratio value close to the nuclear equilibrium value of 4 in most of our objects. These results agree with \citet{palaciosWGB2012} and {\sf AL13}, and are presented in Table~\ref{tab_abund1} and Fig.~\ref{fig:abund}.

In  \citet{palaciosWGB2012}, we reported nitrogen abundances from the literature for four wGb stars (HD  166208, HD 165634, HD 91805, and HD 18474) and
compared them to the predictions of our standard and non-standard stellar evolution models, we concluded that they were compatible with models predictions assuming that these stars were either lower red giant branch stars or red clump stars. In the meantime {\sf AL13} published new abundances for a set of wGb stars derived from high-resolution spectra, revealing a clear nitrogen enrichment of their atmospheres, anti-correlated with the carbon abundance, and interpreted these results as the signature of CN nuclear cycling. Our present study confirms the abundances derived by {\sf AL13}, and indicates strong enrichment (mean value of about $<[N/H]>$ = 1.3 dex) in these stars as shown in Table~\ref{tab_abund1} and Fig.~\ref{fig:abund}.

\begin{table}
\caption{Abundances of heavy elements in wGb stars. All element abundances are given with respect to H, except for lithium
which is given in $\log \epsilon(Li)$. Solar values are taken from \citet{AGS05} and are respectively 7.45, 2.92, and 2.17 for Fe, Sr, and Ba.
}  \label{tab_abund2}
\centering
\begin{tabular}{p{1.3cm}cccccc}
\hline
HD No   & [Fe/H] & eFe & [Sr/H] & eSr & [Ba/H] & eBa  \\         
\hline
\hline
18474 &-0.14 &0.09 &-0.10  & 0.06 & 0.18 & 0.07   \\  
49960 &-0.17 &0.13 & 0.06  & 0.06 & 0.14 & 0.06      \\  
56438 &-0.04 &0.30 & 0.07  & 0.06 & 0.33 & 0.06      \\  
67728 &-0.21 &0.15 & 0.10  & 0.09 & 0.05 & 0.12      \\  
78146 &-0.25 &0.11 & 0.06  & 0.05 & 0.11 & 0.06      \\  
82595 & 0.01 &0.11 & 0.23  & 0.06 & 0.53 & 0.03     \\   
91805 &-0.10 &0.08 & 0.10  & 0.09 & 0.13 & 0.14      \\  
94956 &-0.10 &0.11&   0.09  & 0.07 & 0.39 & 0.03     \\  
102851 &-0.18 &0.12&   0.17  & 0.06 & 0.10 & 0.07     \\  
119256 &-0.06 &0.10&   0.12  & 0.04 &-0.19 & 0.07     \\  
120170 &-0.24 &0.08&  -0.18  & 0.12 & 0.11 & 0.02     \\  
120213 &-0.25 &0.10&  -0.04  & 0.07 & 0.39 & 0.20     \\  
124721 &-0.04 &0.09&   0.07  & 0.06 & 0.25 & 0.07     \\  
146116 &-0.38 &0.13&  -0.11  & 0.08 & 0.37 & 0.05  \\  
165462 &-0.06 &0.11&  -0.18  & 0.06 & 0.27 & 0.09     \\  
165634 &-0.09 &0.09&  -0.08  & 0.09 & 0.15 & 0.11     \\  
166208 &-0.13 &0.11&  +0.22  & 0.11 & 0.16 & 0.04     \\  
204046 &-0.11 &0.10&  -0.08  & 0.03 & 0.14 & 0.12     \\  
207774 &-0.25 &0.11&  -0.12  & 0.07 &-0.02 & 0.03  \\  
 \hline 
\end{tabular} 
\end{table}

\subsection{Errors}\label{sec:errors}
\subsubsection{Random and systematic errors}
Table~\ref{tab_AP} shows the random errors for the basic fundamental
parameters. Errors on effective temperature, microturbulence, and surface
gravity have been derived by evaluating the error due to line-to-line
Fe abundance scatter on  the excitation and equivalent width
trends as well as on ionisation balance, respectively. The resulting errors on those
parameters have then been propagated on the determination of the
luminosity.\\
Because all the stars have similar parameters, systematic errors on
abundances have been evaluated by deriving the abundances in only one
star in our sample and by adding $+1\sigma$ error on stellar
parameters. The resulting values are presented in
Table~\ref{tab:Comparison_abund} (column 2). While they can be almost neglected for C and
Sr abundances, they appear to be significant for the other elements
abundances compared to random errors (see Table~\ref{tab:Comparison_abund} column 4).

\begin{table}
\caption{Errors on abundance determination. Column 2: systematic errors; Column 3 : {\em See notes below}.} %Column 4 : random errors.  }
\centering
\begin{tabular}{ccc}
\hline
[X/H]    & $\Delta$([X/H])\tablefootmark{a} & $\Delta$([X/H])\tablefootmark{b} \\
(dex) & (dex) & (dex)\\
\hline
\hline
C &   0.067  & 0.129 \\
N &   0.134 & 0.116 \\
O &  -0.168 & -0.028 \\
Li &  0.212 & -0.023\\
Na &  0.186 & -0.069 \\
Ba & -0.189 & -0.071\\
Sr &  0.043 & 0.033\\
\hline
\end{tabular} 
\tablefoot{
\tablefoottext{a}{Comparison between abundances deduced from two different
atmosphere models ($T_{\rm eff}=5206~K$, $\log g=2.4$ and $T_{\rm
eff}=5037~K$, $\log g=2.75$) for HD\,56438. For each element X, $\Delta
$([X/H]) = [X/H]$_\mathrm{(5206K ~model)} - $ [X/H]$_\mathrm{(5037K~
model)}$.}\\
\tablefoottext{b}{Comparison between abundances obtained with a C-deficient
atmosphere model and a solar atmosphere model. For each element X,
$\Delta$([X/H]) = [X/H]$_\mathrm{C-deficient~ model} - $ [X/H]$_
\mathrm{solar ~model}$ }
}
\label{tab:Comparison_abund}        
\end{table}   

\subsubsection{Comparison with previous data}
%\textcolor{green}{
Adamczak \& Lambert (2013) have published astrophysical parameters and abundances of 24 wGb stars, of which 13 are in common with our study. Comparison between our work and theirs is given in Table~\ref{tab_AP_comp}.
Our spectroscopic temperatures are consistently higher by 140 $K$ (standard deviation 59 $K$) than those of {\sf AL13}, which results in an overall small discrepancy between our astrophysical parameters and
abundances and theirs, though this discrepancy remains within the limit of respective probable errors.
Nevertheless, there are three
stars (HD\,18474,
HD\,67728, and HD\,120170) for which there are significant discrepancies $\ga 0.3$ dex in abundances; these stars
correspond to those with $\Delta(T_\mathrm{eff}) \ga 200$K in Table~\ref{tab_AP_comp}. 
%}

\begin{table*}[ht]
\caption{Comparison of atmospheric parameters and abundances derived for wGb stars
common to our sample and that of {\sf AL13}. In this table $\Delta$(T) = $T_\mathrm{this work} - T_\mathrm{Adamczak} $ and $\Delta$([X/H]) = $[X/H]_\mathrm{this work} - [X/H]_\mathrm{Adamczak} $
} 
\centering
\begin{tabular}{lrrrrrr}
\hline
HD No   &  $\Delta$(T) & $\Delta$(log g) & $\Delta([Fe/H])$ & $\Delta([C/H])$ & $\Delta([N/H])$  & $\Delta([O/H])$ \\         
& (K) & (dex) \\         
\hline
\hline
HD 18474  &238  & 0.5 & 0.19 &0.30  & 0.27 &  0.38 \\  
HD 49960  &170  & 0.2& 0.08 &0.16  &-0.02 &  0.16 \\  
HD 67728  &197  & 1.1& 0.22 &0.40  & 0.41 &  0.64  \\  
HD 78146  &54  &  0.0 &-0.19 &-0.26 & 0.16 & -0.14\\  
HD 82595  &115  &-0.2 &-0.05 &-0.25 & 0.06 & -0.16\\   
HD 94956  &166 &  0.3 & 0.09 &0.19  &-0.02 &  0.08\\  
HD120170  &237  & 0.4 &0.27 &0.26  & 0.20 &  0.21\\  
HD146116  &120  &-0.1 &0.01 &0.35  &-0.07 & -0.03 \\  
HD165462  &123  & 0.0 &0.02 &0.03  & 0.30 & -0.11\\  
HD165634  &154  & 0.1 &0.09 &0.06  & 0.22 &  0.12\\  
HD166208  &97  &  0.2&-0.13&0.02  & 0.51 & -0.07 \\  
HD204046  &64  &  0.0 &-0.06&-0.08 &-0.03 & -0.01\\  
HD207774  &100  & 0.2 &0.08 &-0.03 & 0.14 &  0.01\\  
 \hline
\end{tabular} 
\label{tab_AP_comp}        
\end{table*}        

 In \S~\ref{sec:APs} we  note a discrepancy between our spectroscopic and photometric temperatures.
Adamczak \& Lambert (2013) report a similar difference of about 190 $K$ between the two techniques without being able to
 clarify this point. We suspect that this is partly due to the chemical peculiarity that affects both
 types of analysis; we discuss this point below.\\

\subsubsection{Influence of the C and N peculiar abundances on
atmospheric parameters and abundances}\label{sec:CNpec}
Because wGb stars have a distinct chemical composition compared to other field
stars, it is expected that this changes the structure of the stellar atmosphere. In this section we study
more quantitatively the impact of the chemical peculiarities on
atmospheric parameters and abundances. 

%% Figure 4
\begin{figure}
\begin{center}
\includegraphics[width=0.37\textwidth,angle=-90]{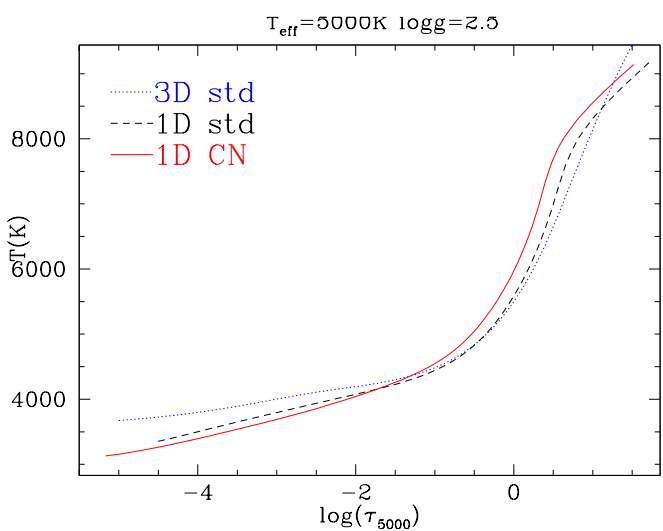}
\caption{Thermal structure comparison between a standard composition
model atmosphere and a model with depleted C and enhanced N. A standard
composition 3D horizontal average model is also displayed
\citep{Magic2013}. }
\label{fig:atmmodels}
\end{center}
\end{figure}
%% Figure 4

We compute two model atmospheres with the MARCS code
\citep{Gustafsson2008}, both with the same parameters chosen to represent a typical wGb
star in our sample ($\rm T_{eff} = 5000~K, \log g = 2.5~{\rm dex}, [Fe/H] = 0~{\rm dex}$),  one with standard (e.g. solar) C and N composition ($\rm[C/H] = 0.0~ dex, [N/H] = 0.0 ~dex $) and the other with the composition of a wGb star ($\rm[C/H]=-1.30~dex , [N/H]=+1.20~dex$).\\

Fig.~\ref{fig:atmmodels} shows the difference in the thermal structure
between the two models. While in the outer layers (at small optical depths) the CN-specific model
is slightly cooler than  the standard composition model, the inner part is warmer at the continuum formation level. 

\begin{figure}
\begin{center}
\includegraphics[width=0.37\textwidth,angle=-90]{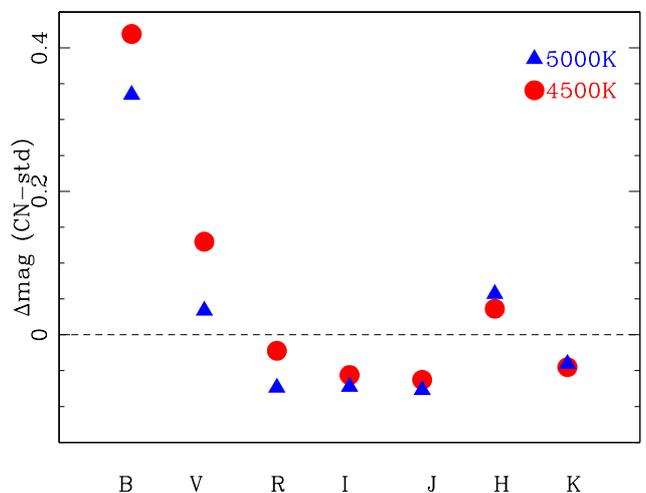}
\caption{Difference in magnitude of the Johnson filters when using
CN-specific models and standard composition model.}
\label{photom}
\end{center}
\end{figure}

Because the atmosphere structure is significantly affected, one can also
expect that the spectral energy distribution and the
photometric magnitudes will also be modified. In Fig.~\ref{photom}, we show the difference in magnitude obtained with CN-specific models compared to standard composition
models with \teff = 5000 K and 4500K. Not surprisingly, the differences are larger for the B and V
bands because this is where most of the C-based and N-based molecules
are the most prominent. The typical differences in colours are $\Delta (B-V) =
+0.3$ , $\Delta (V-I) = +0.1$, and $\Delta(V-K) = +0.1$, which  translate
 in an increase of the photometric effective temperature of 400 $K$, 200 $K$,
and 100$K$, respectively, when accounting for the CN peculiar abundances in the model atmosphere computation. This is consistent with the values expected from the thermal structure. \\For the spectroscopic temperature, we were also able to evaluate
the difference in stellar parameters obtained with both kinds of models
by measuring the change in the excitation trend and the change in
ionisation balance for HD~56438. We obtained an increase of 150$K$ and 0.5 dex
for  \teff and $\log g$ when using the proper CN model. The change in $\log g$ implies a decrease in the luminosity of -0.5 in log scale (see Eq.~\ref{logL}).\\

Accounting for the peculiar abundances of wGb stars when performing the spectral
synthesis would generally lead to identifying the wGb stars as less evolved and
less massive stars. 
%% Figure 5
%% Figure 5

Using a chemically dedicated model atmosphere to analyse wGb stars
affects positively the effective temperature derived by photometry and
by spectroscopy (but not necessarily by the same amount).  However, we
cannot claim straightforwardly that this effect would concile the
discrepancy in the parameters observed because this would require a
fully consistent analysis for the whole sample with a complete grid of
dedicated model atmospheres.

 In Table~ \ref{tab:Comparison_abund}
(column 3)  we report the difference in
abundances according to the type of model atmospheres used. While it appears negligible for all the elements measured
through atomic transitions, accounting for the CN peculiarity in the model atmopshere and spectra synthesis does affect the C and N
abundances. Very interestingly, we note that the
positive change in both C and N abundances indicates that wGb stars could actually be
even more enhanced in C+N than  is found based on solar scaled model atmospheres.\\

Despite the evidence that  the chemical peculiarity of wGb
stars has some impact on the parameters and the determination of the abundances, it is also important to
simultaneously consider 3D effects on the structure of the atmosphere. Indeed,
for similar stellar parameters, 3D models  show non-negligible
differences against 1D standard composition models (Fig.~
\ref{fig:atmmodels}), but  hydrodynamical models with both low C and high
N composition  do not exist at the moment, and thus we are currently
bound to 1D standard composition models. 

However, we emphasize that the overall change in parameters
with appropriate C and N opacities, as we demonstrate in this section,
suggests that wGb
stars are probably hotter and with a higher surface gravity, which
implies that they could be even less evolved than  we infer in \S~
\ref{sec:evol}.

\section{Dynamical analysis of the sample}\label{sec:kinematics}

In order to complete our study, we  analysed the kinematics of wGb stars in our sample. To this end, we calculated the Galactic space velocity components (U, V, and W) given the radial velocities derived from our data and the proper motions and parallaxes according to \cite{Hipparcos07}. We use a right-handed coordinate system for U, V, and W so that they are positive in the direction of the Galactic center, Galactic rotation, and the north Galactic pole respectively \citep{JS87}.
We  corrected for a solar motion of $(U_{\odot}, V_{\odot},W_{\odot})=(8.5; 13.38; 6.49)$~km.s$^{-1}$ \citep{Coskunoglu2011}.

%% Figure 6
\begin{figure*}[ht]
\begin{center}
%\vspace*{-1.7cm}
\includegraphics[width=0.3\textwidth]{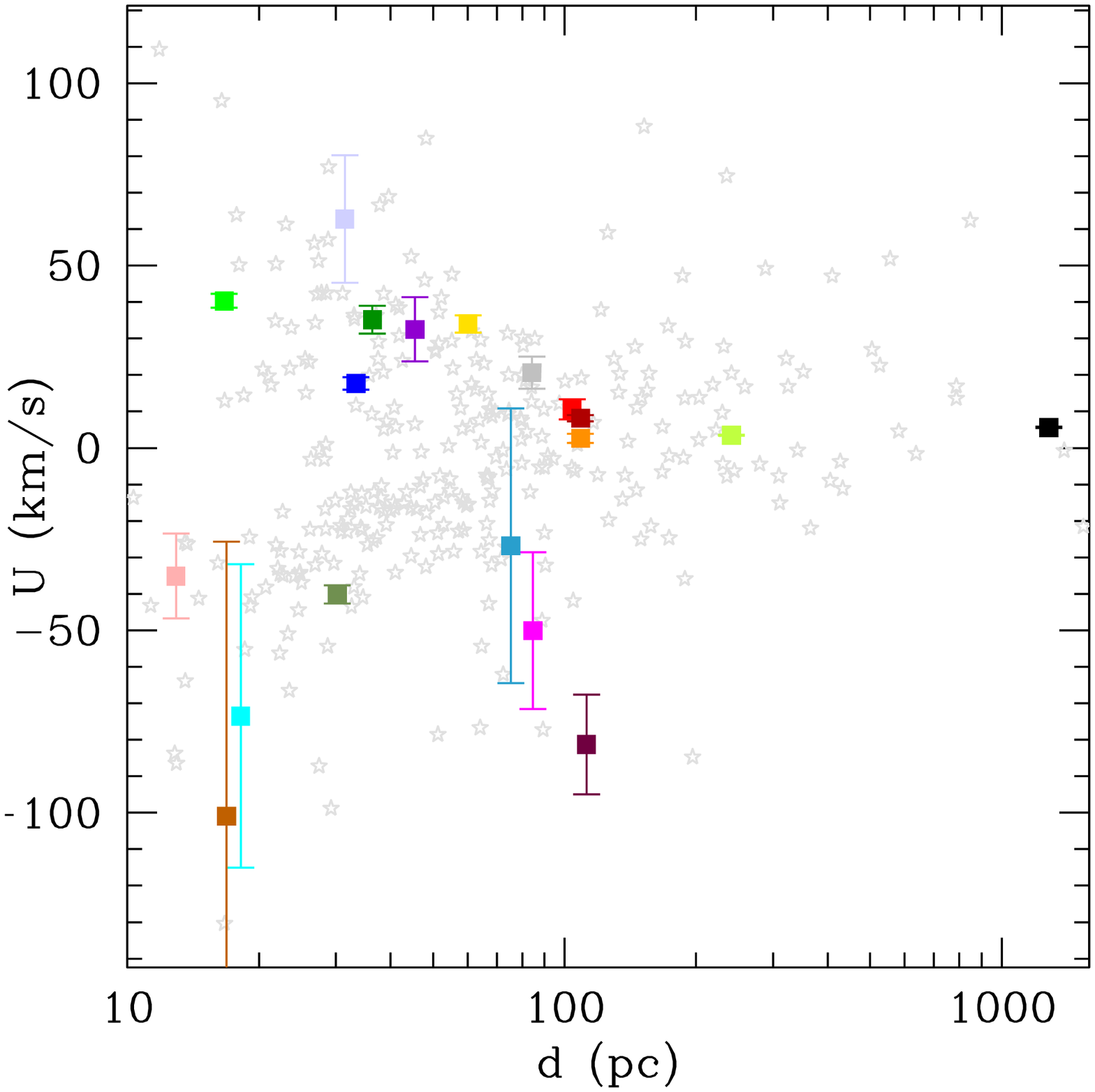}%
\includegraphics[width=0.3\textwidth]{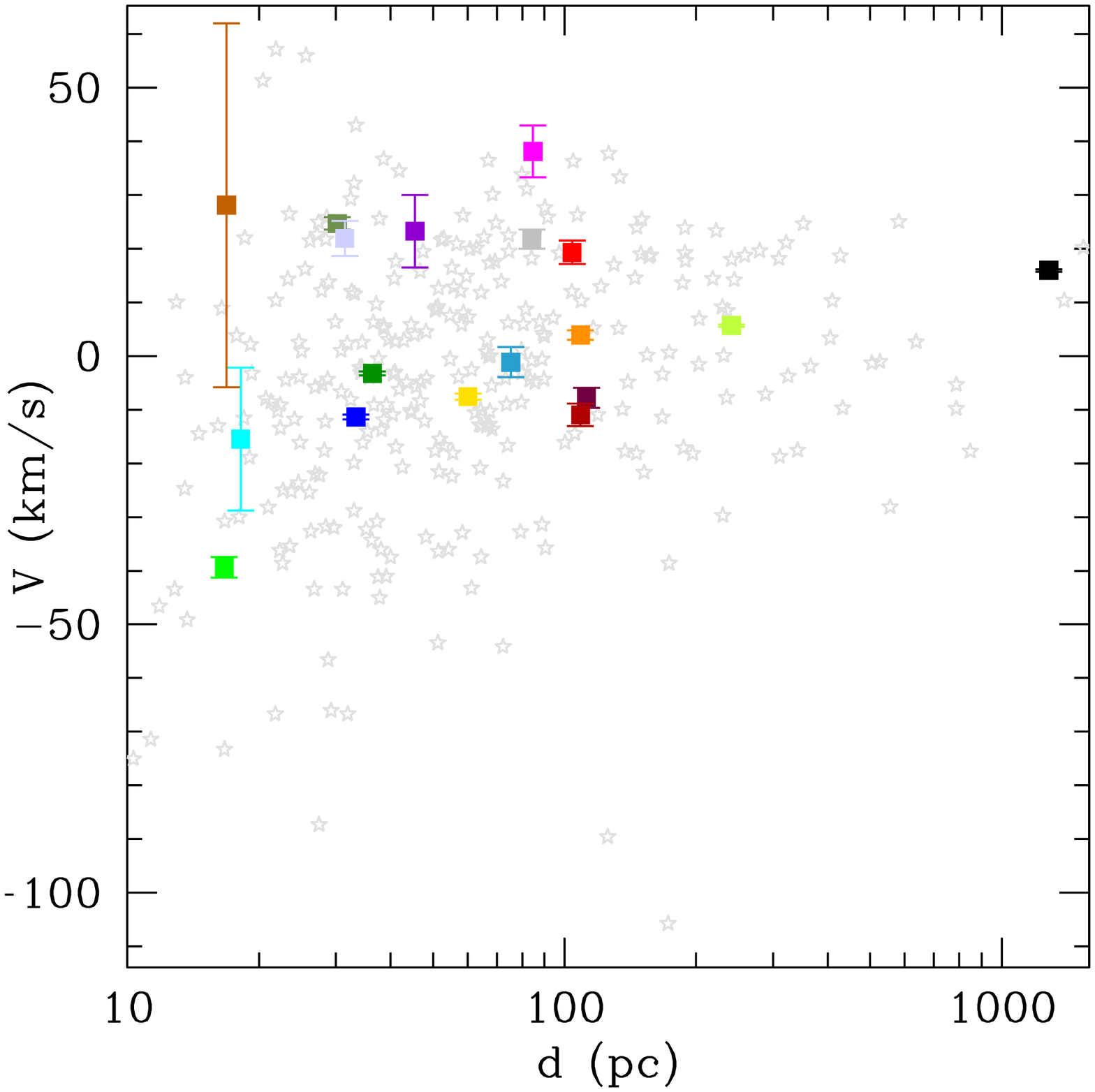}%
\includegraphics[width=0.3\textwidth]{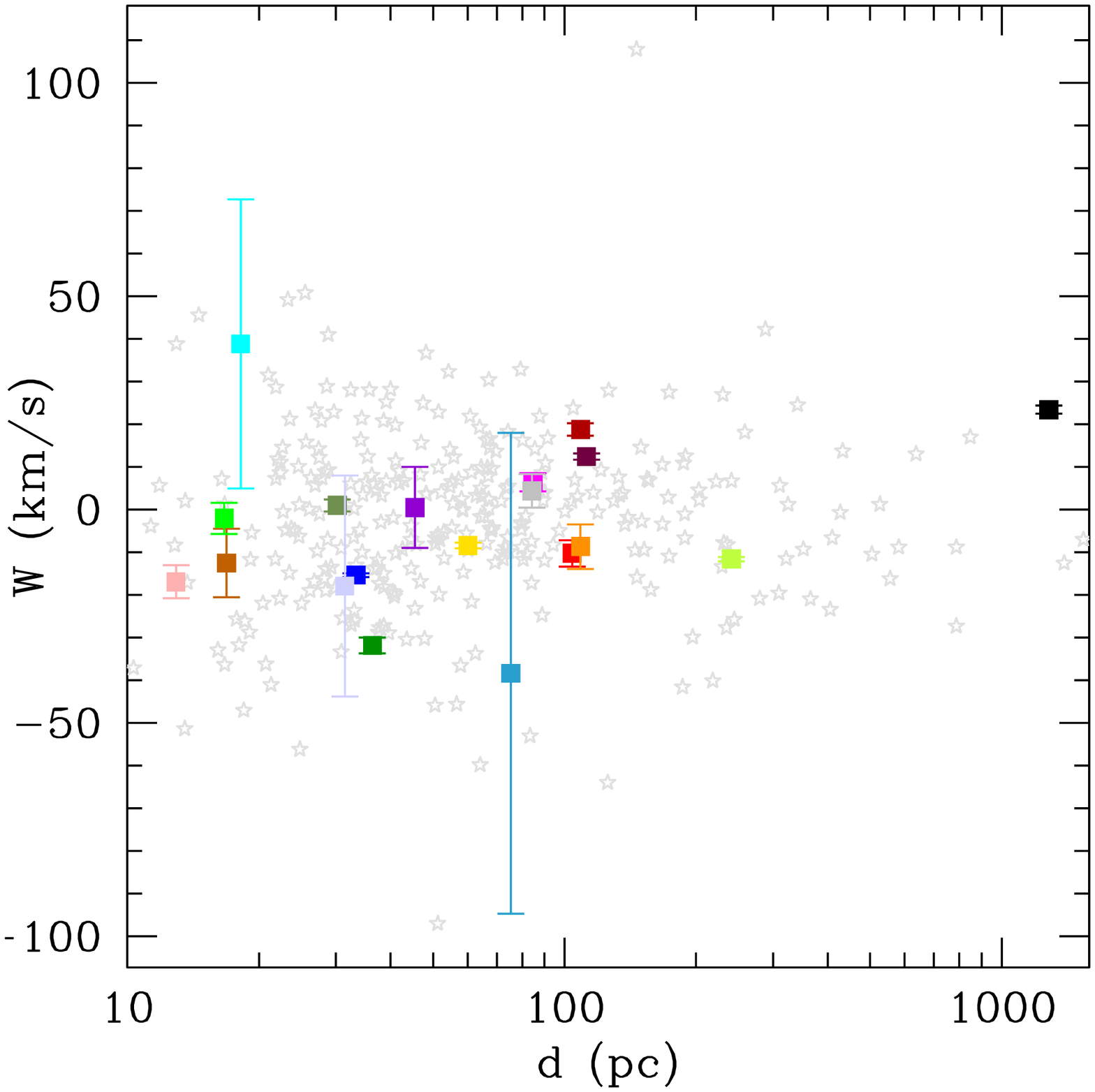}
\includegraphics[width=0.3\textwidth]{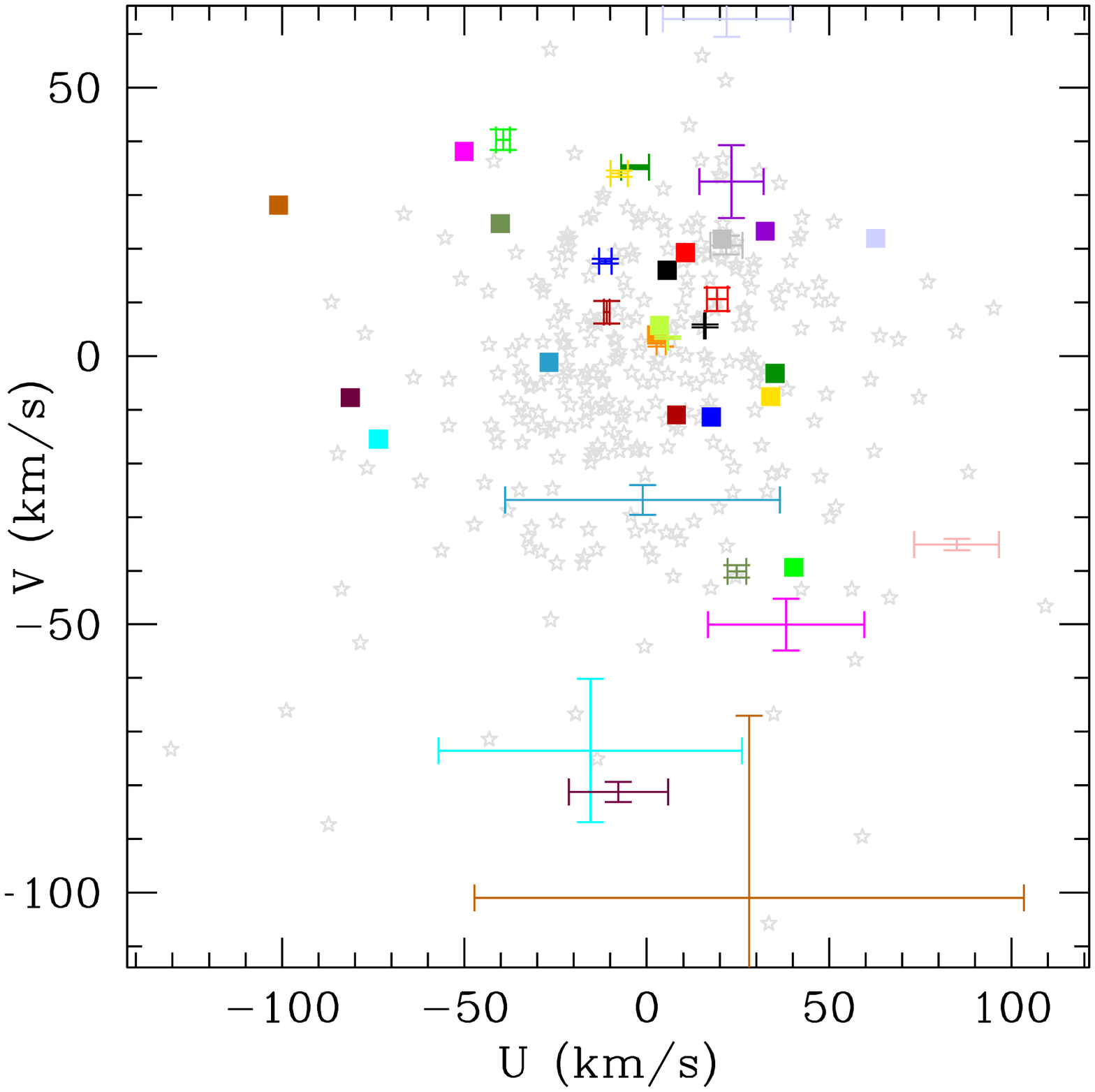}
\includegraphics[width=0.3\textwidth]{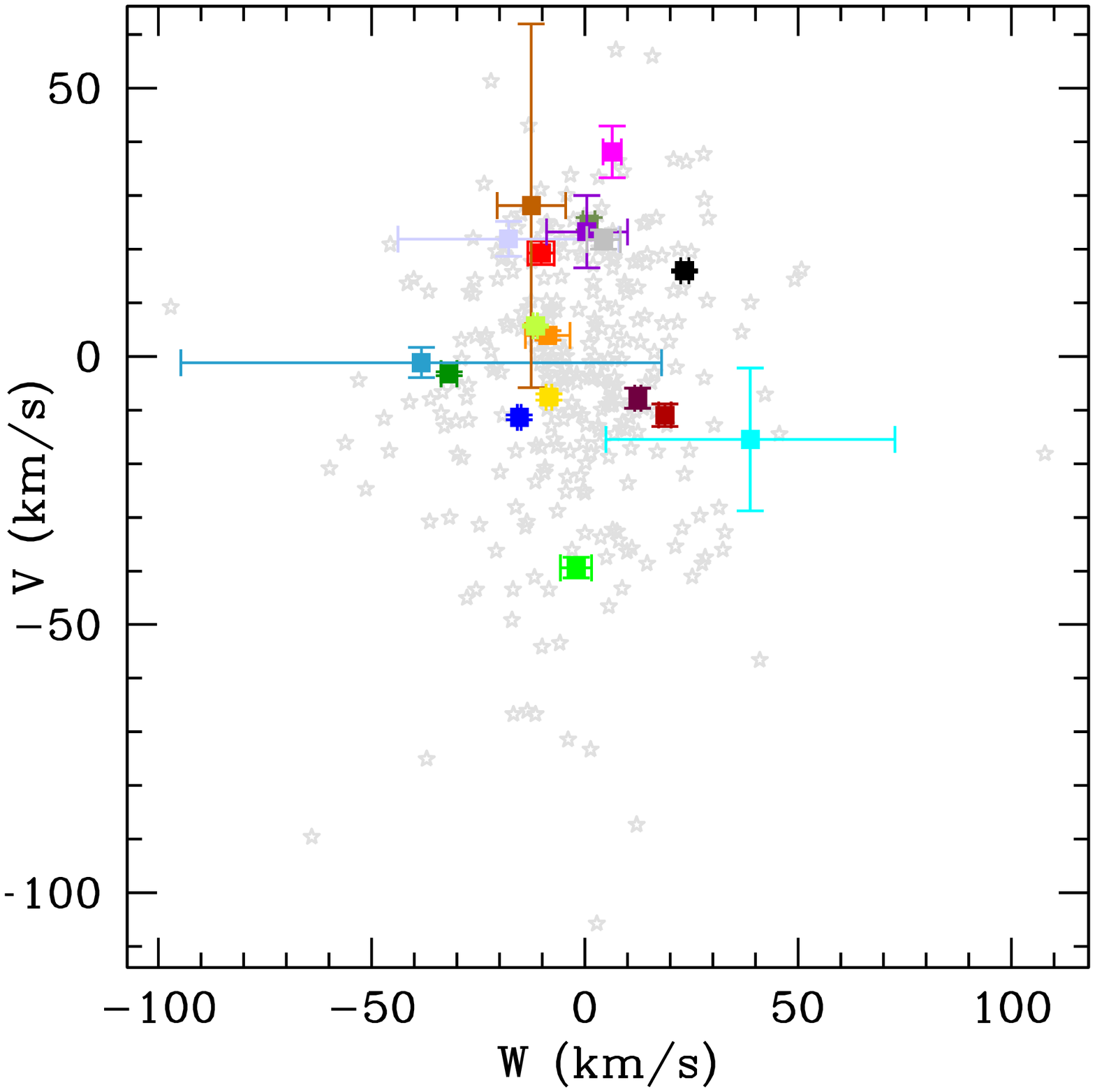}%
\includegraphics[width=0.3\textwidth]{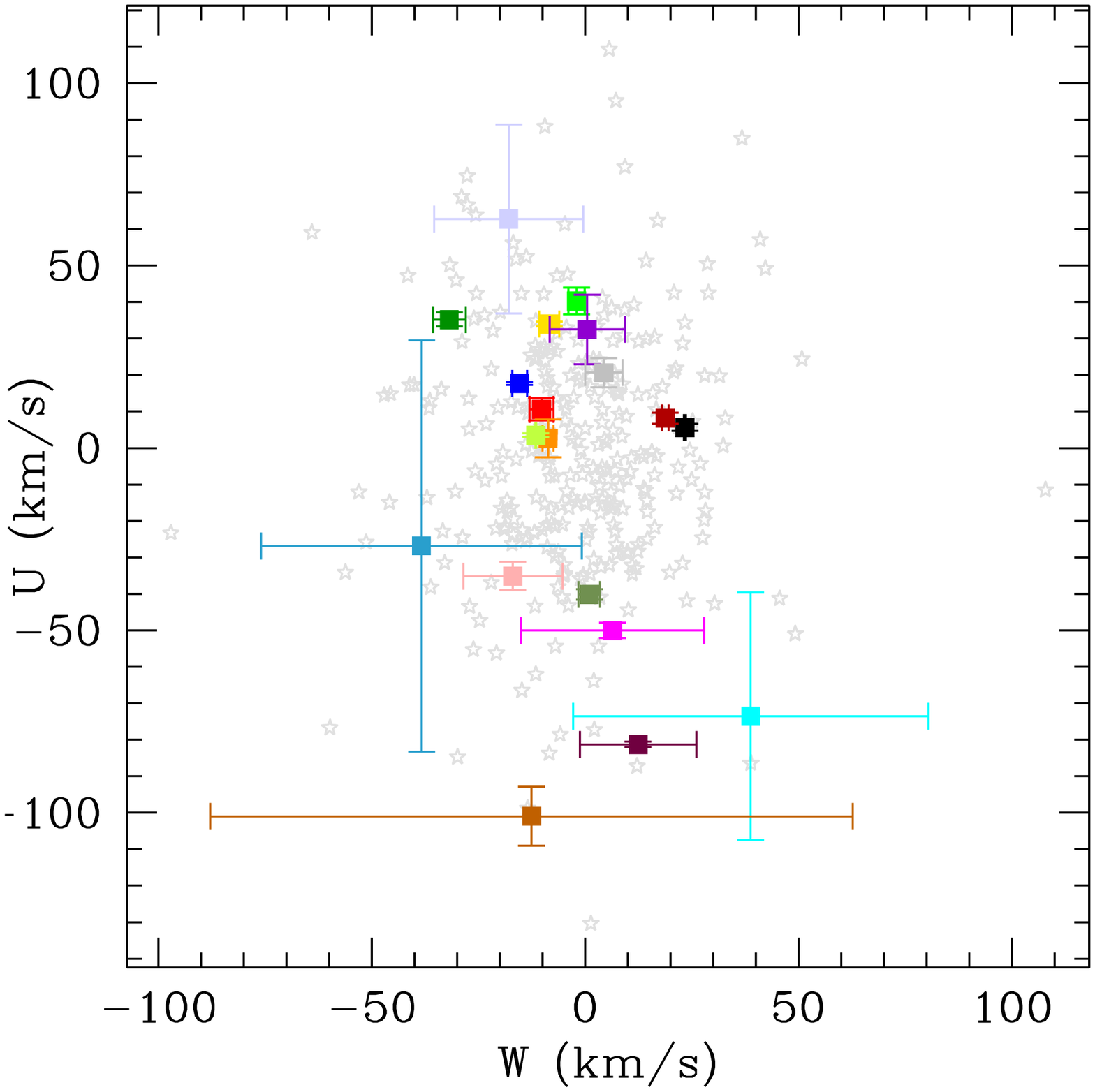}\\
%\vspace*{1cm}
\end{center}
\caption{UVW positions for the sample stars (according to colour-coding in Table~\ref{tab_AP}) and the Galactic red giants sample by \citet{LH07} (grey stars in the background). The upper row gives the velocity components as a function of the distance to the Sun and 
the lower row shows the velocity components plotted against each other.}
\label{fig:kin}
\end{figure*}
%% Figure 6

The uncertainties associated with the space-velocity components are obtained from the observational quantities and their errorbars after the prescription of \citet{JS87}. The points bearing the largest uncertainties are those for which the parallax is the most uncertain. In order to provide a point of comparison, we  applied the same procedure to the red giant stars sample of \citet{LH07} to which {\sf AL13} also compared their work.\\
The results are shown in Fig.~\ref{fig:kin}. The kinematics of the wGb stars in our sample do not follow any specific trend and do not differ from those of normal red giant stars in the Galaxy. In this way we exclude the possibility that the chemical peculiarities of these stars  originate from a common birth place and kinematic history.

\section{Comparison to stellar evolution models}\label{sec:evol}

Following a similar approach to the one we used in \citet{palaciosWGB2012}, we have used the STAREVOL v3.30 stellar
evolution code to compute grids of dedicated models to analyse the
data described here. The reader is referred to \citet{siess06,decressin09} for a full description of the standard and non-standard physics of the STAREVOL code used to compute these grids. \\
Consistently with the model atmospheres and
spectral synthesis, we have used the \citet{AGS05} values as a
reference for the solar chemical composition. Unless specified otherwise, the adopted value for
the heavy elements mass fraction is 0.012294 with $\alpha_{\rm MLT}$ =
1.6014 as obtained by the calibration of a solar model. The different grids computed are described in Table~\ref{tab_models}. In addition to a grid of standard models at solar metallicity (labelled {\sf S} in Table~\ref{tab_models}), in which we considered an overshoot (at the edges of all convective regions) of 0.1 H$_{\rm p}$, we computed three grids dedicated to the wGb stars problem. First, we computed a standard grid with [Fe/H] = -0.25 (labelled {\sf SlZ} in Table~\ref{tab_models}) in order to explore the impact on the mass and evolutionary status that can be assigned to the wGb stars in our sample considering that most of them are slightly iron poor. We used the same amount of overshooting as for the solar metallicity models.\\ Assuming that the CNO abundances determined for the wGb stars in our sample could have an extrinsic origin, we also computed a grid setting the initial abundances of these nuclides to the mean value of our observed sample (see Table~\ref{tab_models}, row labelled {\sf CNO}). \\ 

\begin{table*}[t]
\caption{Grids of dedicated stellar evolution models computed from the pre-main sequence (PMS) to the early-AGB stage. All models are computed using \citet{AGS05} as a reference for the solar abundances mixture (Z$_\odot$ = 0.012294).} 
\centering
\begin{tabular}{lcccccccc}
\hline
Type  & $< \upsilon_{ZAMS} > $  & M & Z & [C/Fe]$_{ini}$ & [N/Fe]$_{ini}$  & [O/Fe]$_{ini}$  & [Na/Fe]$_{ini}$ & Mixing \\         
& km/s & (M$_\odot$) &  & (dex) & (dex) &  (dex) &  (dex) &\\         
\hline
\hline
S & 0 & 3.0 - 5.5 & 0.012294 & 0 & 0 & 0 & 0 &  overshoot $\alpha_{ov} = 0.1 H_p$ \\
SlZ & 0 & 3.0 - 5.5 & 0.0070188 & 0 & 0 & 0 & 0 & overshoot $\alpha_{ov} = 0.1 H_p$  \\
CNO & 0 & 3.0 - 5.5 & 0.025824 & -1.218 & 1.295 & 0.238 & 0.370 &overshoot $\alpha_{ov} = 0.1 H_p$ \\
RotCNO & 25 & 3.5 ; 3.8 & 0.0070188 & -1.218 & 1.295 & 0.238 & 0.370 & overshoot $\alpha_{ov} = 0.1 H_p$\\
&&&&&&& & rot. mix. \citet{MZ98}\\
SlowRot & 25 & 3.0 - 5.5 & 0.012294 & 0 & 0 & 0 & 0 & overshoot $\alpha_{ov} = 0.1 H_p$\\
&&&&&&& & rot. mix. \citet{MZ98}\\
FastRot & 180 & 3.0; 3.5 & 0.012294 & 0 & 0 & 0 & 0 & overshoot $\alpha_{ov} = 0.1 H_p$\\
&&&&&&& & rot. mix. \citet{MZ98}\\
%Rotation2 & 200 & 3.0 - 5.5 & 0.012294 & 0 & 0 & 0 & 0 & \citet{MZ98}\\
%Mixing & 25 & 3.0 - 5.5 & 0.012294 & 0 & 0 & 0 & 0 &  \citet{MZ98} + $D_{wGb}$\\
 \hline
\end{tabular} 
\label{tab_models}        
\end{table*}     

As in \citet{palaciosWGB2012}, we also computed models including rotation. We know from \citet{palaciosWGB2012,AL13} that wGb stars have an initial mass between 3 and 5 \msun, which means that their progenitors were late B-type stars on the main sequence. Following the results of \citet{ZR12}, we  computed a series of models in this mass range with an initial slow rotation of $\upsilon_{\rm ZAMS} = 25 $~km.s$^{-1}$, which is representative of 5\% of late B-type stars. These models are referred to as {\sf SlowRot} in Table~\ref{tab_models} and their surface rotation velocity at the Zero Age Main Sequence (ZAMS) is given in Col. 2. Two models at 3.5 and 3.8 M$_\odot$ were computed using the same initial CNO abundances as in the  {\sf CNO} models but with [Fe/H] = -0.25 dex and $\upsilon_{\rm ZAMS} = 25 $~km.s$^{-1}$, which we label {\sf RotCNO}. The bulk of late B-type stars have a mean rotation velocity of about 220 km.s$^{-1}$ on the main sequence, and we computed a few models (M = 3.0, 3.5 \msun) of fast rotators in order to indicate the level of abundance variations expected in that case. They are referred to as {\sf FastRot} in Table~\ref{tab_models} and are discussed in \S~\ref{sec:abund}.

We include rotation through its modification of the gravitational potential by the centrifugal force (following the \cite{ES76} formalism) and via the transport of angular momentum and nuclides that it generates. We use the \citet{MZ98} formalism where the transport of angular momentum is modelled by an advection-diffusion equation. Meridional circulation, turbulent shear, and winds are the processes transporting both angular momentum and nuclides. We  use the  \citet{zahn92} and \citet{TZ97} prescriptions for the horizontal and vertical shear viscosities, respectively. \\ Mass loss is described according to \citet{reimers75} using a parameter $\eta_{Reimers} = 0.5$.\\

\subsection{Masses and evolutionary status}
Using the standard grids at solar and subsolar metallicity we  estimated the masses and radii for all the stars in
our sample following the maximum likelihood method described in
\citet{valle14}. We  based our estimates upon the \teff, \logg,
and [Fe/H] and associated errorbars derived from our spectroscopic
analysis. The derived values and associated 1$\sigma$ errors are given in Table~\ref{tab_MRage}. We note that the masses obtained when using modified ab initio CNO abundances should be larger by about 0.2 M$_\odot$, since varying the C and N abundances while keeping the same iron content leads to an overall  increase in the metallicity and thus to a shift towards redder less luminous tracks at a given mass. They still remain in good agreement with the values listed in Table~\ref{tab_MRage} within the 1$\sigma$ errorbars. 

For those stars common to both our sample and that of {\sf AL13}, the errorbars and the mass domain that we derive are smaller. This is due to a combination of the difference of our \teff by $\approx 140$ K and  the use of different stellar evolution models and methods to derive the masses. A striking example of the differences that we find is the star HD 120170. Adamczak \& Lambert (2013) list it as a 0.6 M$_\odot$ star, while we identify it as a 3.6 M$_\odot$ in good agreement with its temperature and gravity, which are  inconsistent with a 0.6 M$_\odot$ track. For this specific star, the large uncertainty on the parallax deduced luminosity can be incriminated to explain the low value found by {\sf AL13}.

\begin{table}
\caption{Masses and radii of program stars} 
\centering
\begin{tabular}{lcccc}
\hline
HD No   &  M &  $\sigma$ & R & $\sigma$ \\         
& (M$_\odot$) & & (R$_\odot$) &  \\         
\hline
\hline
HD18474       &3.2   &-0.2 / + 0.6  & 14.3   & -3.0 / +3.6     \\
HD49960       &3.2   &-0.2 / +0.6  & 14.2   & -2.4  / +5.3    \\
HD56438       &3.1   &-0.1 / +0.1  & 12.25   & -1.15  / +1.55   \\
HD67728       &3.8   &-0.65 / +0.6  & 23.2   & -9.1  / +16.3   \\
HD78146       &3.9  &-0.5 / +0.1  & 28.2   & -6.1  / +7.5   \\
HD82595       &3.7   &-0.2 / +0.2  & 22.5   & -2.1  / +4.0   \\
HD91805       &3.2   &-0.1 / +0.5  & 15.9   & -3.4  / +4.7  \\
HD94956       &3.3   &-0.3 / +0.7  & 12.9   &-6.1  / +12  \\
HD102851      &3.2   &-0.15 / +0.8  & 13.7   & -4.7  / +12.5  \\
HD119256      &3.2   &-0.2 / +0.8  & 14.4   & -3.9  / +6  \\
HD120170      &3.2   &-0.2 / +0.8  & 12.1   & -1.9  / +4.5  \\
HD120213      &4.2   &-1.2 / +0.7  & 34.3   & -14  / +17.9   \\
HD124721      &3.3   &-0.2 / +0.7  & 14.4   & -3.7  / +7.5  \\
HD146116      &4.0   &-0.7 / +0.2  & 31.5   & -6.7  / +3.4  \\
HD165462      &3.5   &-0.4 / +0.5  & 17.8   & -5.8  / +11.9   \\
HD165634      &3.3  &-0.2 / +0.6  & 16.4   & -5.4  / +7.1  \\
HD166208      &3.2   &-0.2 / +0.8  & 11.3   & -1.1  / +3.7   \\
HD204046      &3.4   &-0.3 / +0.8  & 16.4   & -7.5  / +17.8   \\
HD207774      &3.4   &-0.4 / +0.6  & 13.2   & -6.5  / +13.8  \\
 \hline
\end{tabular} 
\label{tab_MRage}        
\end{table}  
Concerning the evolutionary status, the location of our sample stars in the HR diagram is not constraining as shown in Figures~\ref{fig:HRLiSGB} and~\ref{fig:HRLiclump}. The temperature and luminosity of the stars are all compatible with a subgiant / red giant evolutionary status, whichever  model is considered (Fig.~\ref{fig:HRLiSGB}). On the other hand, the extent of the blue loops (corresponding to the core He burning  or clump phase) is very sensitive to the opacity in the outer layers of the models, and can be largely modified by a change in the chemical composition of these layers as illustrated in  Fig.~\ref{fig:HRLiclump}. In particular the hotter and less luminous wGb stars in our sample are not compatible with a clump evolutionary status. The blue loops are much shorter in the rotating models ({\sf SlowRot} and {\sf FastRot}) and in the models with modified CNO abundances, and only four stars are compatible with a clump status from their location in the HR diagram when compared to the {\sf CNO} and {\sf FastRot} sets of models.\\
According to the discussion in \S~\ref{sec:errors}, adopting the \teff from {\sf AL13} for the stars common to our samples would shift most of the points towards the red in Figures~\ref{fig:HRLiSGB} and \ref{fig:HRLiclump}, making them more compatible with a clump status. On the other hand, adopting fundamental parameters derived from dedicated model atmospheres including the CN peculiarity of the wGb stars would shift most of the points towards the blue and would almost exclude a clump status.

As is clearly seen here, the location in the HR diagram alone cannot be conclusive in this particular region for intermediate-mass stars and there is a clear degeneracy between red giant and core He burning phases. Asteroseismology based on space photometry is to date the only method that can clearly disentangle stars from the clump and stars from the red giant branch \citep{Bedding11}, but unfortunately none of the known wGb stars has been observed by the asteroseismic space missions yet.

In order to gain insight on the evolutionary status of wGb stars,  other observational constraints can be added, such as lithium abundance. This dimension is added as a colour code both for the tracks and for the observational points in Figs.~\ref{fig:HRLiSGB} and ~\ref{fig:HRLiclump}. It is clear from these figures that all stars with A(Li)$ > 1.3$ dex are not compatible with a post first dredge-up (1st DUP) completion status based on stellar evolution models predictions. The completion of the 1st DUP corresponds in Fig.~\ref{fig:HRLiSGB} to the point with the highest temperature at which the lowest lithium abundance is reached in each of the tracks.  About half of our sample presents such lithium abundances (9 out of 19). According to our computations, lowering the initial metallicity and/or introducing  rotation leads to even lower lithium abundances at the 1st DUP completion and hence at the clump (A(Li) $< 0.95$ dex), and there are even more wGb stars (12 out of 19 within the quoted errorbars) that exhibit more lithium at their surface than predicted by the models.\\ Rotational velocity may also be added to the information on temperature, luminosity, and surface lithium abundance to gain insight on the evolutionary status of wGb stars. However, the upper limits obtained on the surface velocities \vsini~for all but two of the wGb stars studied here are not restrictive enough to help determine their evolutionary status. All our rotating models are compatible with \vsini$ < 5$ km.s$^{-1}$ both during the Hertzsprung gap and at the clump as shown in Fig.~\ref{fig:rotvsT}.\\

From an evolutionary point of view, the time spent by a star at the  clump (i.e. core He burning phase) is much longer than that spent in the Hertzsprung gap and low red giant branch (RGB) ($<$ $\Delta $t $> _{SGB/RGB} \approx$ 1.1 to 3 Myr vs. $<$ $\Delta$ t $> _{clump} \approx$ 14 to 155 Myr in the mass range 3 to 5 \msun , with lower durations associated with higher masses), which is usually an argument in  favour of  a clump status for stars for which the only constraint is the location in the area of the HR diagram where clump and subgiant/red giant tracks are intertwined.\\ In the specific case of the very rare wGb stars (less than 0.3\% of the G-K giants), although no clear statement can be made, their rarity combined with their lithium content, rotational velocities, and positions in the HR diagram seem to favour a mildly evolved status. A clump status is, however, not fully ruled out and could be attributed to at least four objects in our sample. If the wGb stars  cross the Hertzsprung gap,  important questions arise, in particular concerning the whereabouts of progenitors and the descendants of wGb stars, which we  address in \S~\ref{sec:scenario}.

\subsection{Nucleosynthesis and abundance patterns}\label{sec:abund}

\begin{figure*}
\begin{center}
%\vspace*{-2cm}
\includegraphics[width=0.4\textwidth]{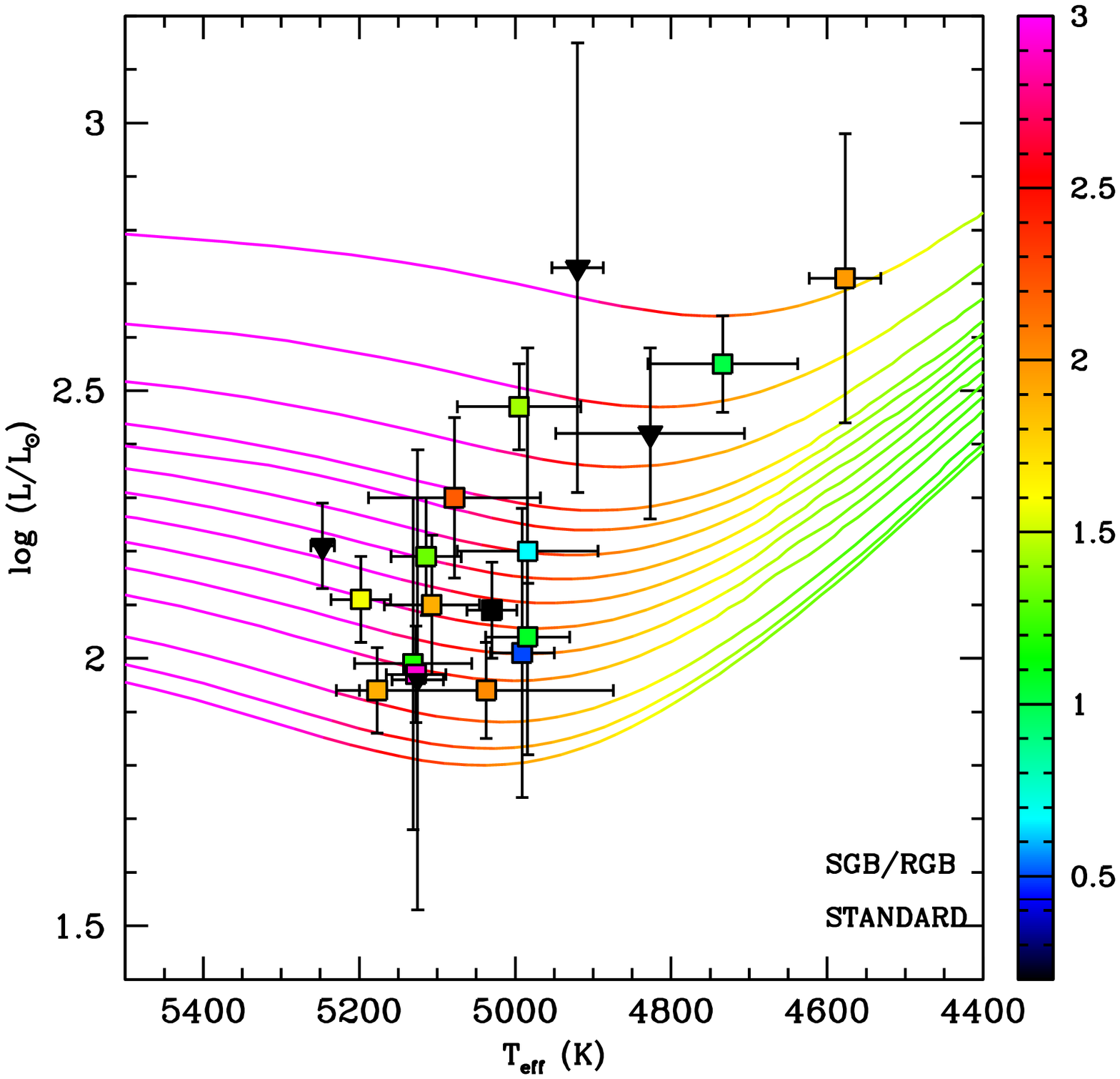}%
\includegraphics[width=0.4\textwidth]{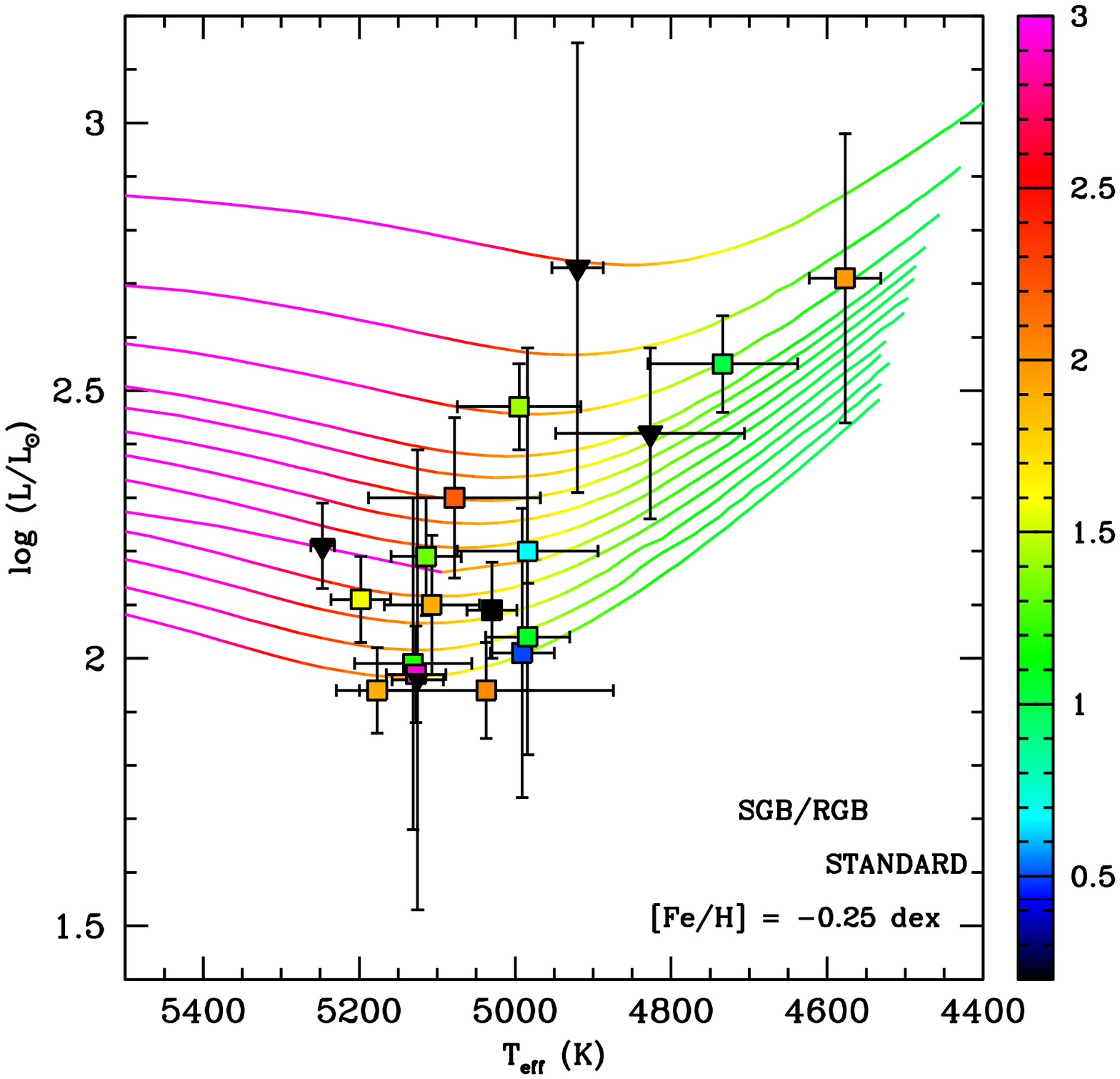}
\includegraphics[width=0.4\textwidth]{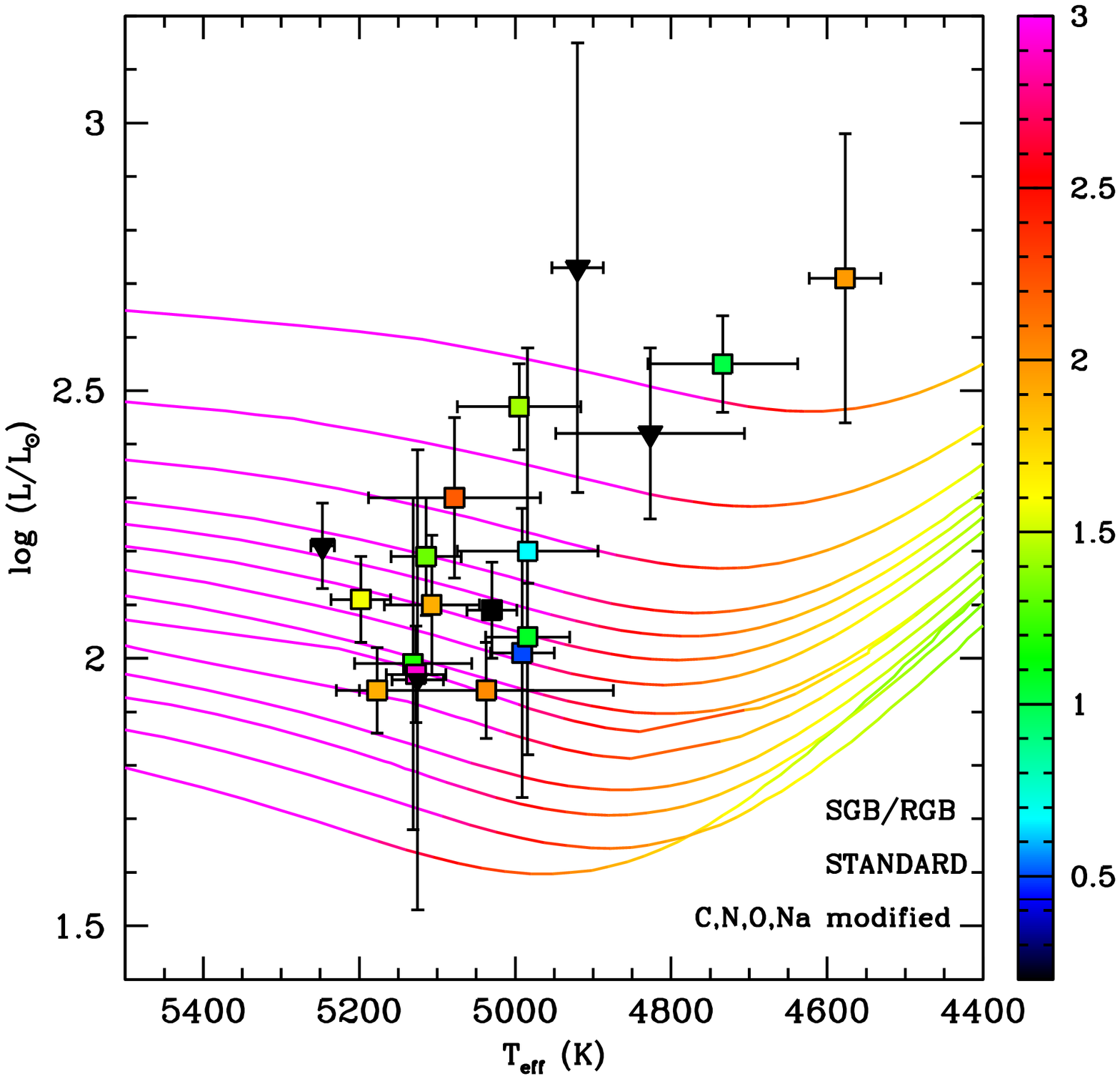}%
\includegraphics[width=0.4\textwidth]{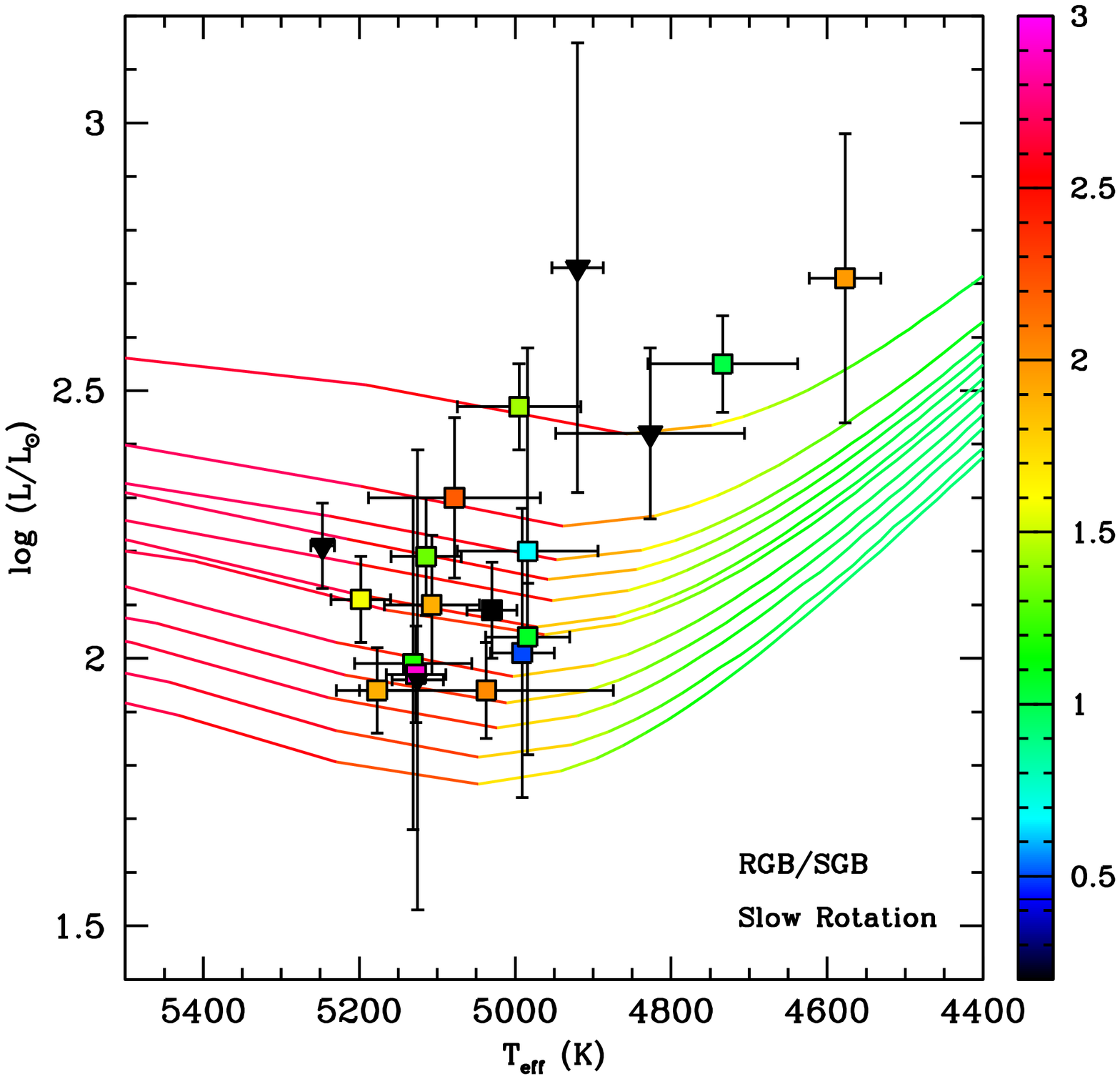}
\includegraphics[width=0.4\textwidth]{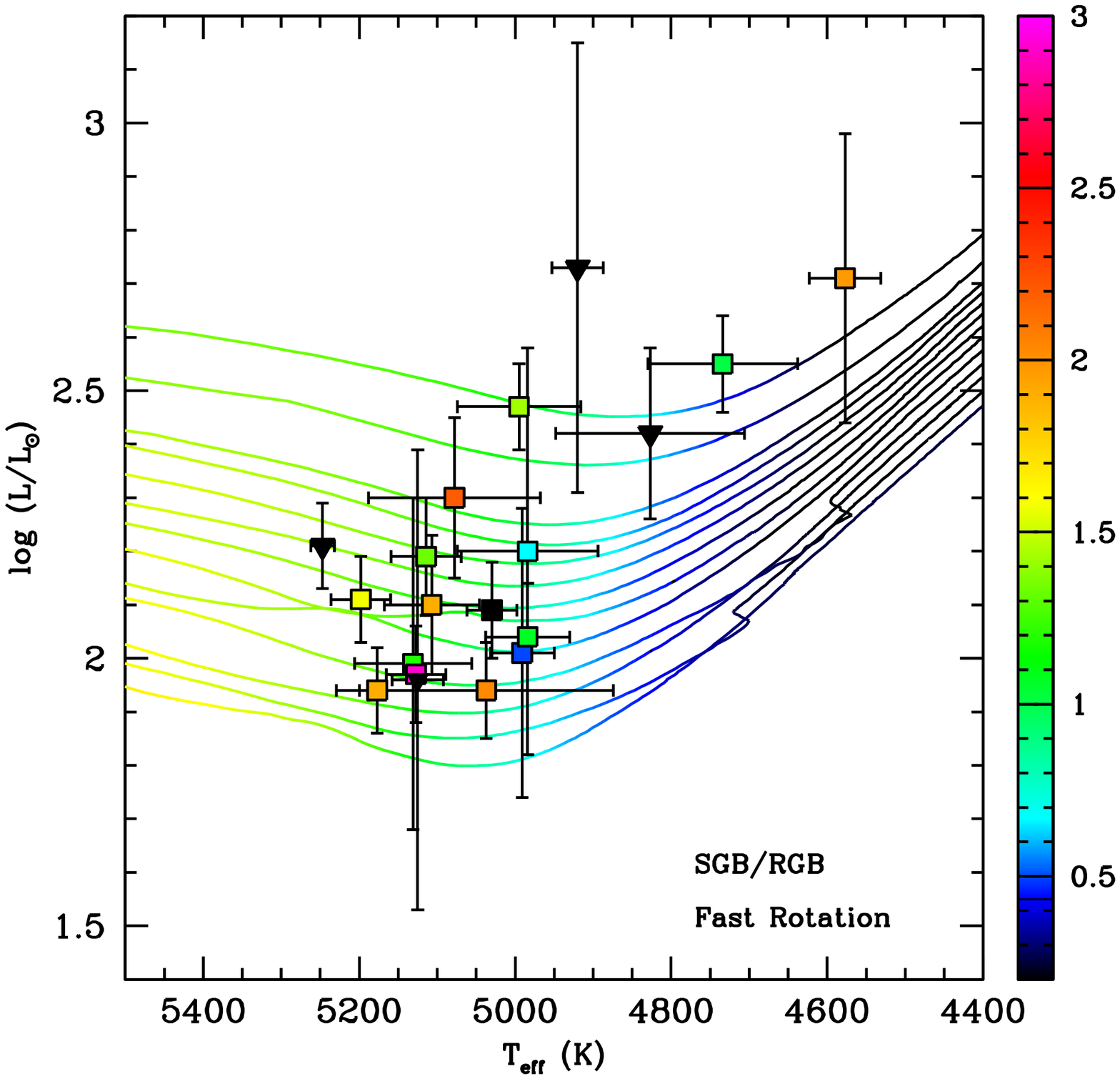}
\end{center}
%\vspace*{1.6cm}
\caption{Hertzsprung-Russell diagrams for the models of the different grids described in Table~\ref{tab_models}. Only the part of the tracks associated with the subgiant and red giant phases is plotted here. The colour code of the tracks indicates the surface abundance of lithium (see colour scale on plots). Our data points are overplotted with the same lithium abundance colour-coding  used for the evolutionary tracks. The tracks correspond to models with initial masses between 3 M$_\odot$ (least luminous track) and 5 M$_\odot$ (most luminous track).}
\label{fig:HRLiSGB}
\end{figure*}

\begin{figure*}
\begin{center}
%\vspace*{-2cm}
\includegraphics[width=0.4\textwidth]{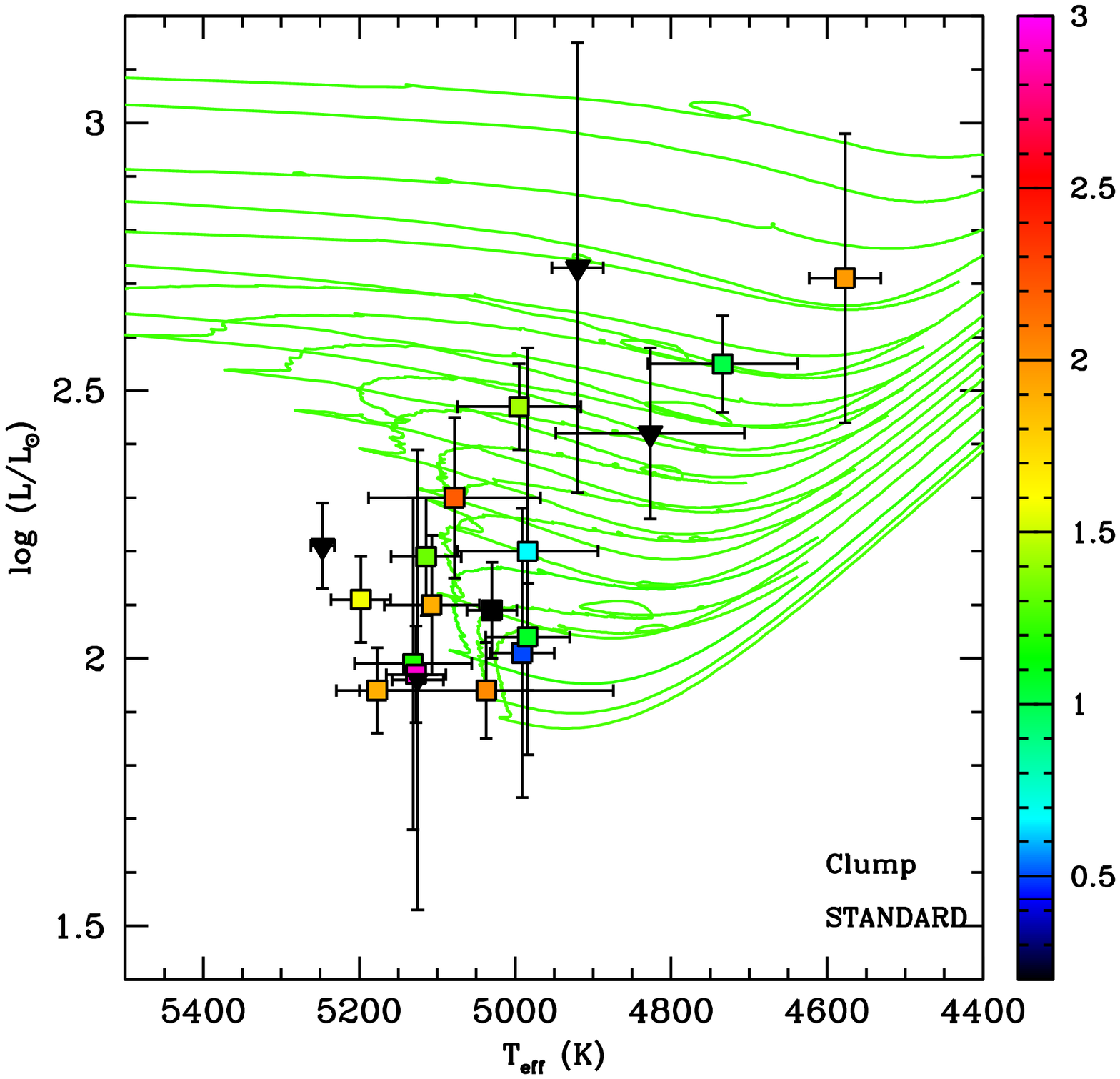}%
\includegraphics[width=0.4\textwidth]{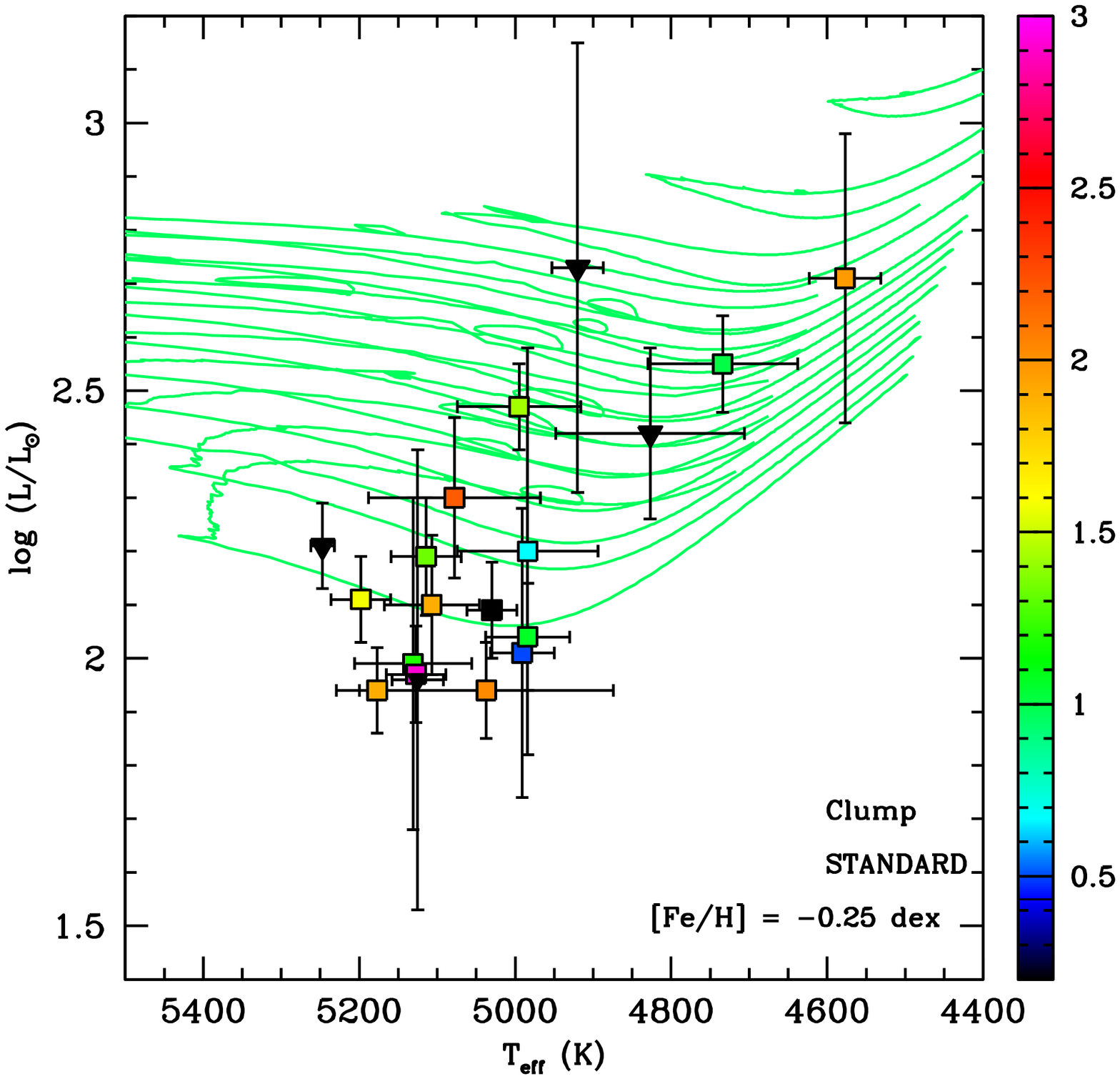}
\includegraphics[width=0.4\textwidth]{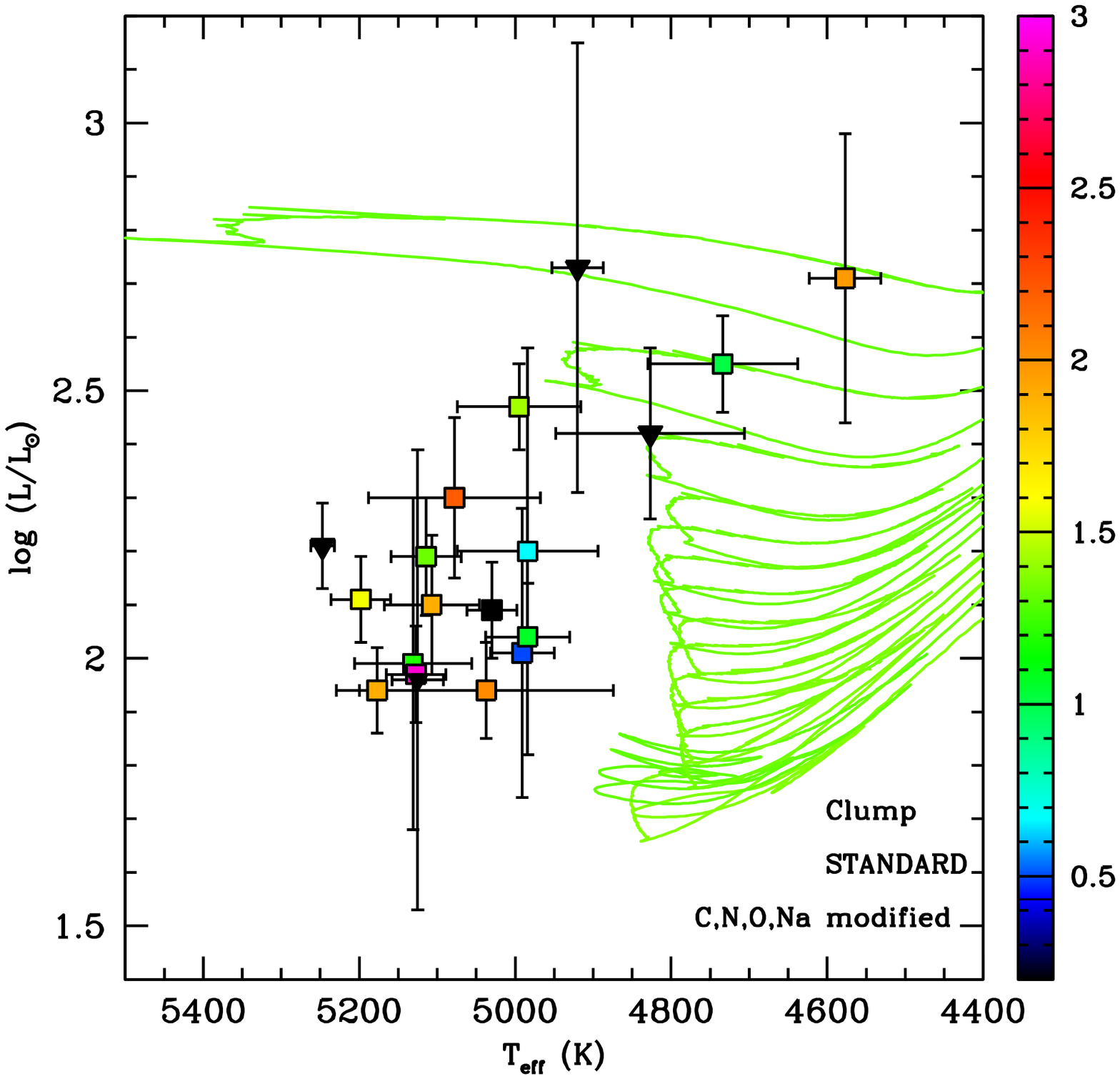}%
\includegraphics[width=0.4\textwidth]{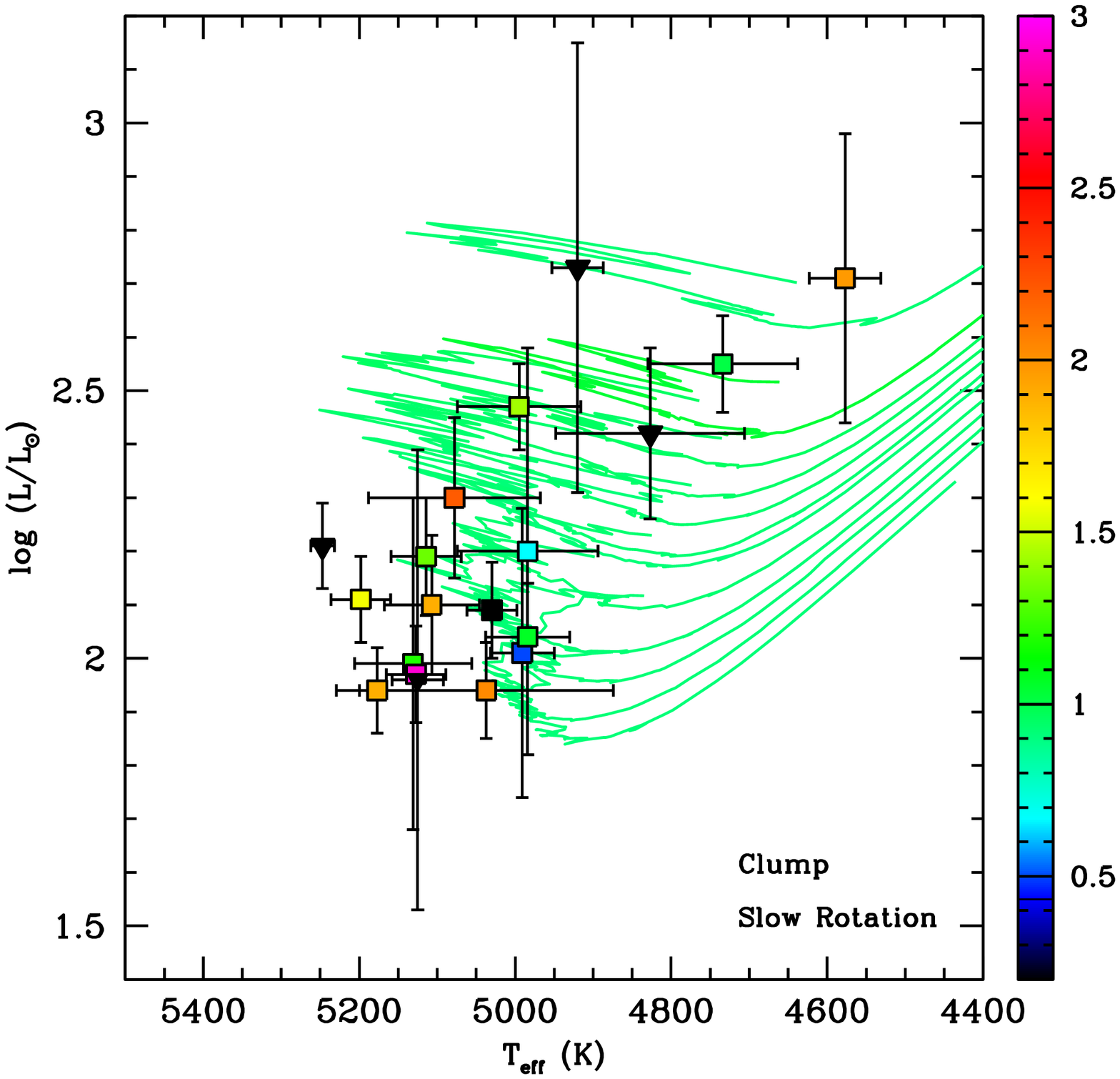}
\includegraphics[width=0.4\textwidth]{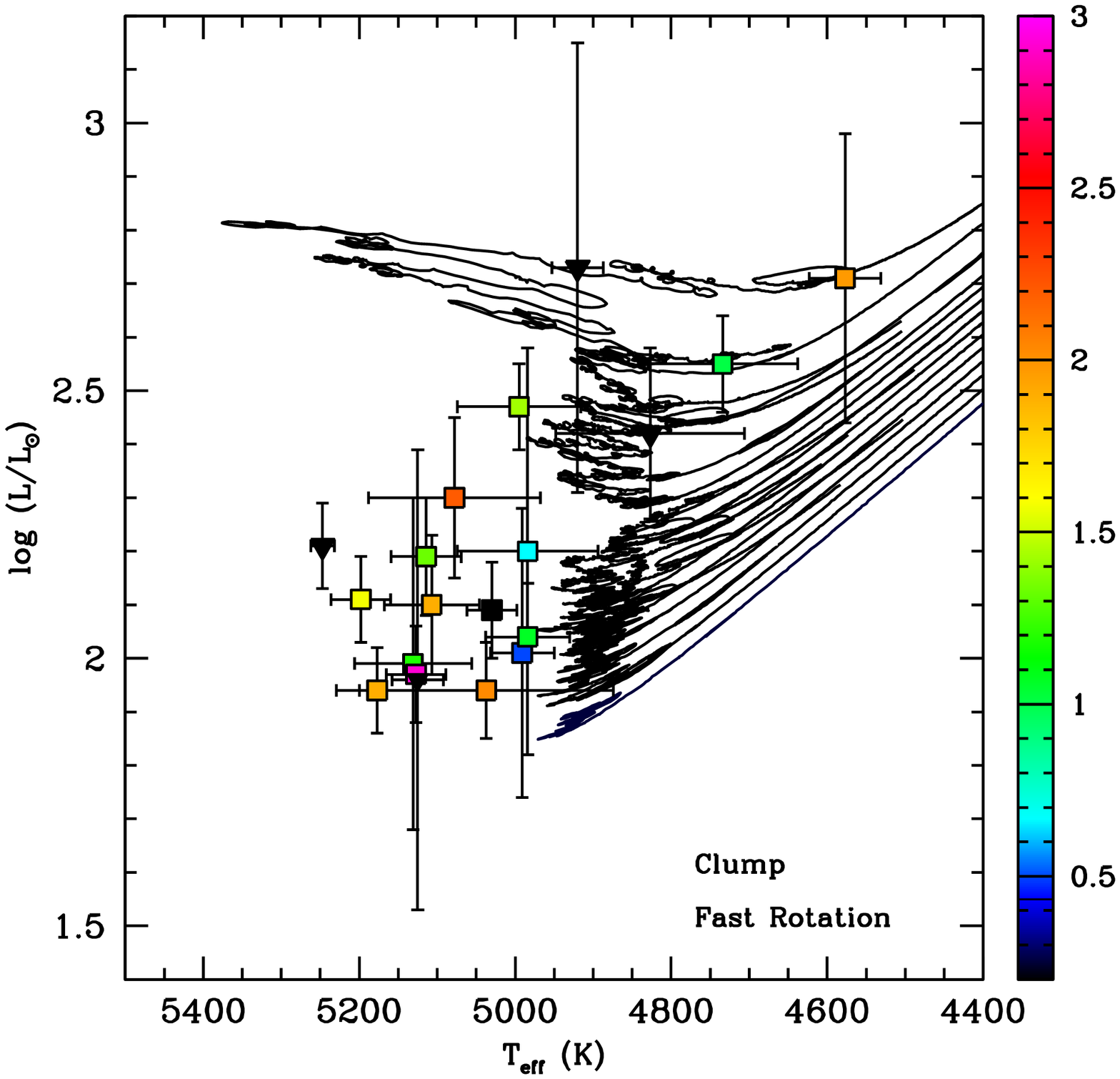}
\end{center}
%\vspace*{1.5cm}
\caption{Same as Fig.~\ref{fig:HRLiSGB} for the part of the tracks associated with the core helium burning phase (clump).} 
\label{fig:HRLiclump}
\end{figure*}

The expected evolution of the surface abundances of lithium, carbon, and nitrogen as a function of the effective temperature predicted by the different models of intermediate-mass stars described in Table~\ref{tab_models}, is shown in Figures~\ref{fig:HRLiSGB}, ~\ref{fig:HRLiclump}, and~\ref{fig:abund}. 
In Fig.~\ref{fig:abund}, the colours of the points are the ones listed in the last column of Table~\ref{tab_models}, and refer to each individual wGb star in our sample.

\subsubsection*{Lithium}

Lithium is easily destroyed by proton captures and its abundance decreases at the surface as soon as the base of the convective envelope is connected to regions where T $\geq 2.5 \cdot 10^6$ K. In all our models this occurs during the 1st DUP with the deepening of the convective envelope. The lithium abundance log $\varepsilon(Li) \equiv A_{\rm Li}$ then decreases at the surface by about 2 dex compared to the value at the turn-off, reaching $A_{\rm Li} \approx 1.2$ dex in standard models.\\ In rotating models, the surface lithium has already decreased during the main sequence owing to the efficient turbulent transport occurring in the radiative interior and connecting the base of the convective envelope to the inner regions where lithium is destroyed by proton captures. During the 1st DUP, $A_{\rm Li}$ also drops by 2 orders of magnitude, starting at the turn-off from initial values between 2.5 dex and 1.5 dex for $\upsilon_{ZAMS}$ between 25 and 50 \kms.\\

%% Figure 7
%% Figure 7

%% Figure 8
%% Figure 8
\clearpage

At least 9 out of the 19 wGb stars analysed here exhibit lithium abundances higher than the larger expected post DUP value of 1.2 dex. These stars are most probably young giants crossing the Hertzsprung gap with on-going 1st DUP.
The remaining stars (10 out of 19) have lithium abundances between 0 and 1.2 dex that are consistent with 1st DUP dilution in rotating models, and thus compatible with both subgiant and clump status.

%% Figure 9
\begin{figure}
\begin{center}
%\vspace{-2.1cm}
\includegraphics[width=0.45\textwidth]{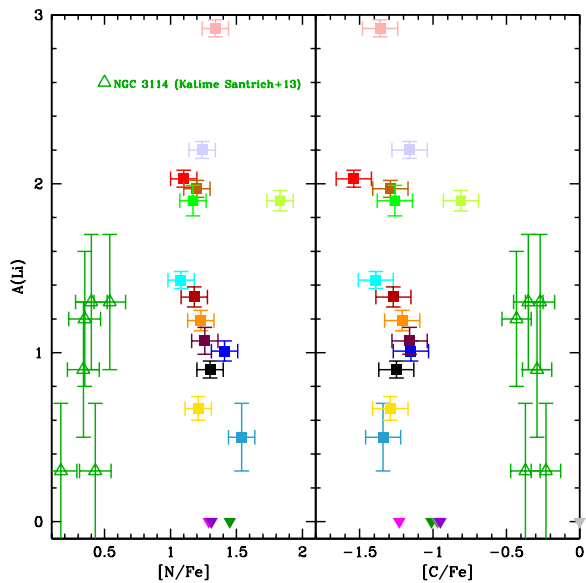}
\end{center}
%\vspace{1.8cm}
\caption{Derived Li abundances as a function of [N/Fe] and [C/Fe] for the wGb stars in this study (coloured squares and inverted triangles according to the colour code in Table~\ref{tab_AP}) and from the open cluster NGC 3144 after \citet{Santrich13} (green 
triangles).}
\label{fig:LiNC}
\end{figure}
%% Figure 9

Fig.~\ref{fig:LiNC} presents the lithium abundances as a function of the C and N abundances for the wGb stars and for intermediate-mass stars in the same range of \teff and luminosities from the NGC 3144 young open cluster study by \citet{Santrich13}. No correlation exists between the Li, C, and N abundances in wGb stars. On the other hand, the stars in NGC 3144 are all Li-poor, with abundances consistent with evolved rotating post-dredge up stars. Thus the wGb stars do not behave as their siblings from the NGC 3144 cluster, which have been shown by \citet{Santrich13} to be similar to field counterparts as far as C and N abundances are concerned.

%% Figure 10
\begin{figure}
\begin{center}
%\vspace{-2.1cm}
\includegraphics[width=0.45\textwidth]{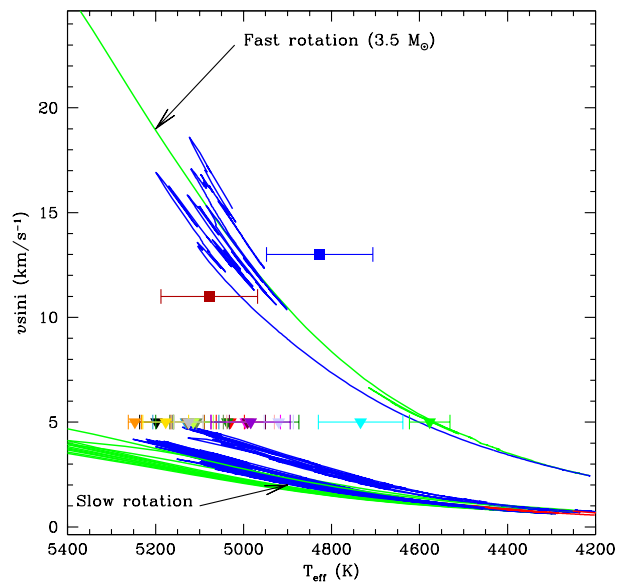}
%\vspace{1.8cm}
\caption{\vsini~as a function of temperature for our solar metallicity slow rotating models with masses from 3 to 5 M$_\odot$ and for a fast rotator of 3.5 M$_\odot$. The colour of the data points refers to each individual wGb star in our sample as specified in Table~\ref{tab_AP}. The colour of the tracks indicates the evolutionary phase: green for subgaint and RGB phases, blue for clump.}
\label{fig:rotvsT}
\end{center}
\end{figure}
%% Figure 10

\subsubsection*{Carbon}

\begin{figure*}[t]
\begin{center}
%\vspace*{-2cm}
\includegraphics[width=0.45\textwidth]{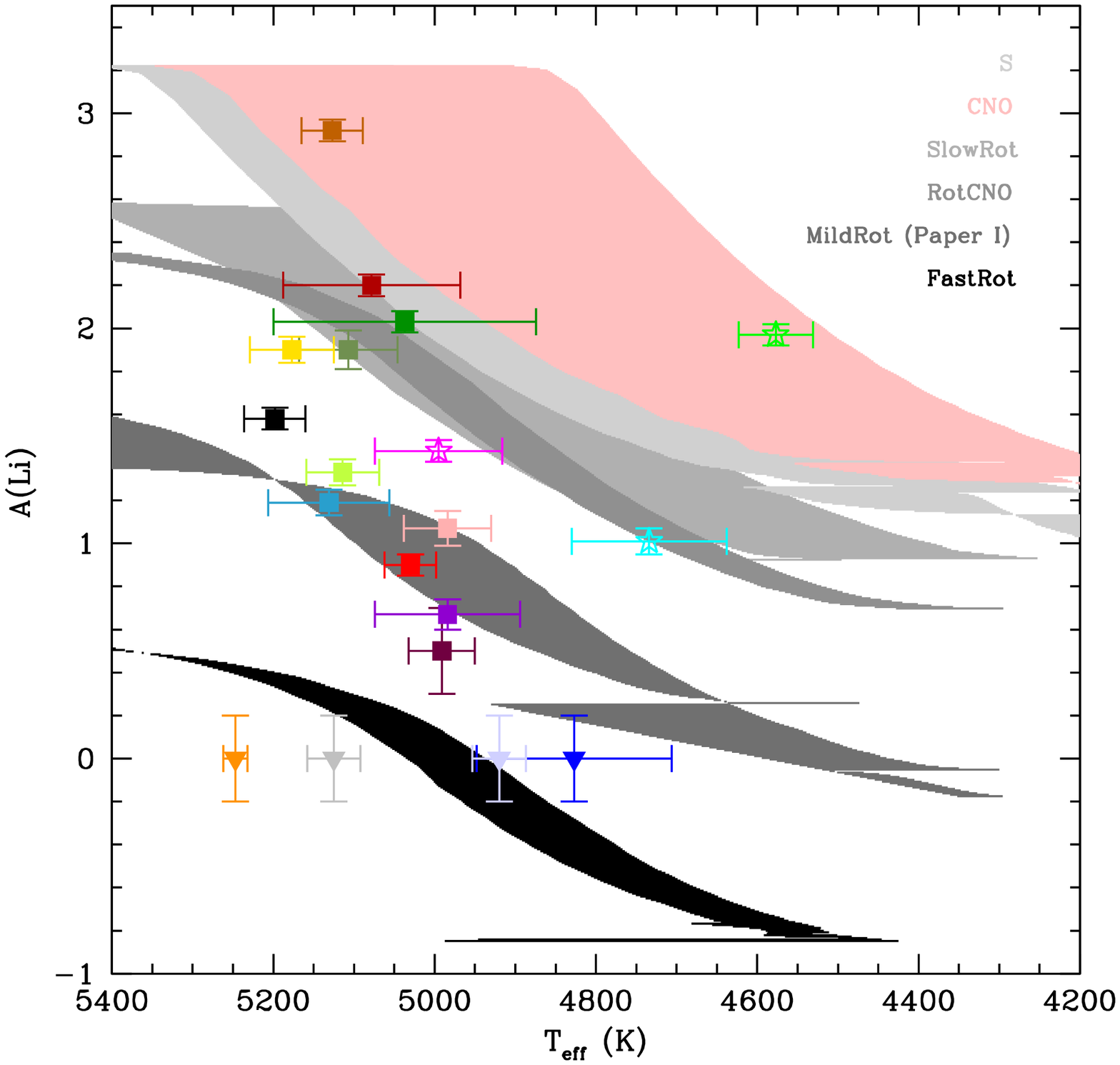}%
\includegraphics[width=0.45\textwidth]{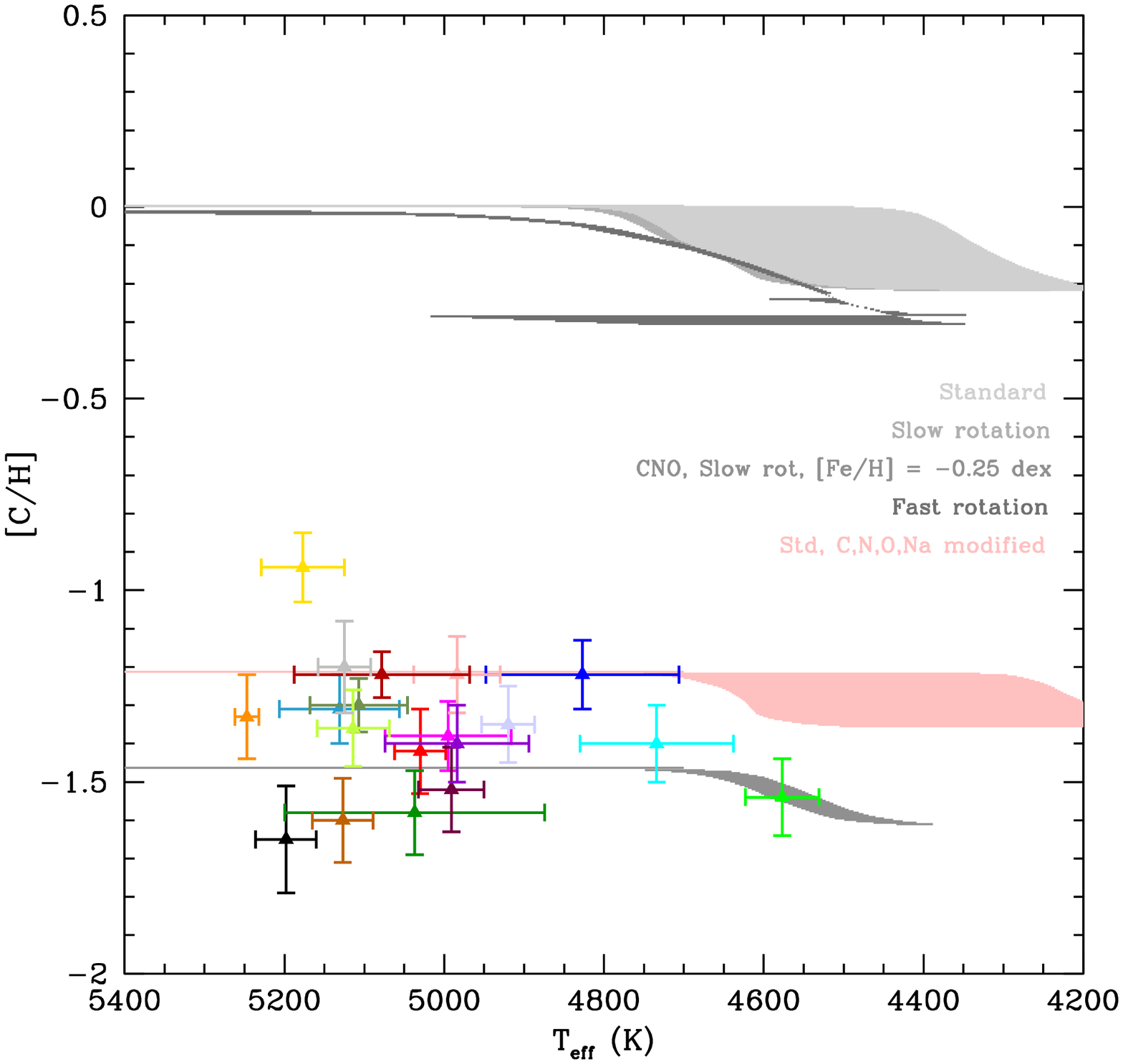}
\includegraphics[width=0.45\textwidth]{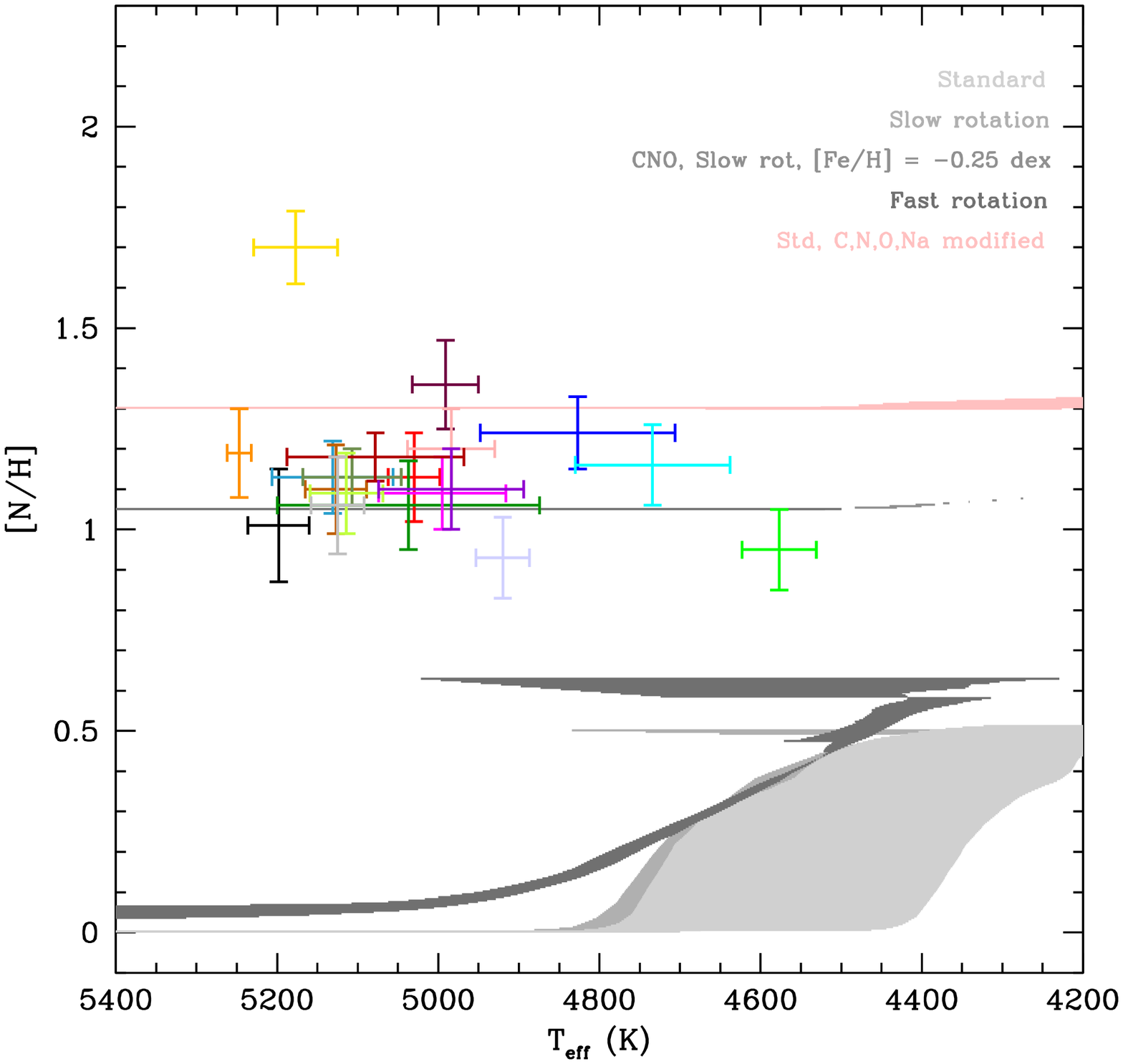}%
\includegraphics[width=0.45\textwidth]{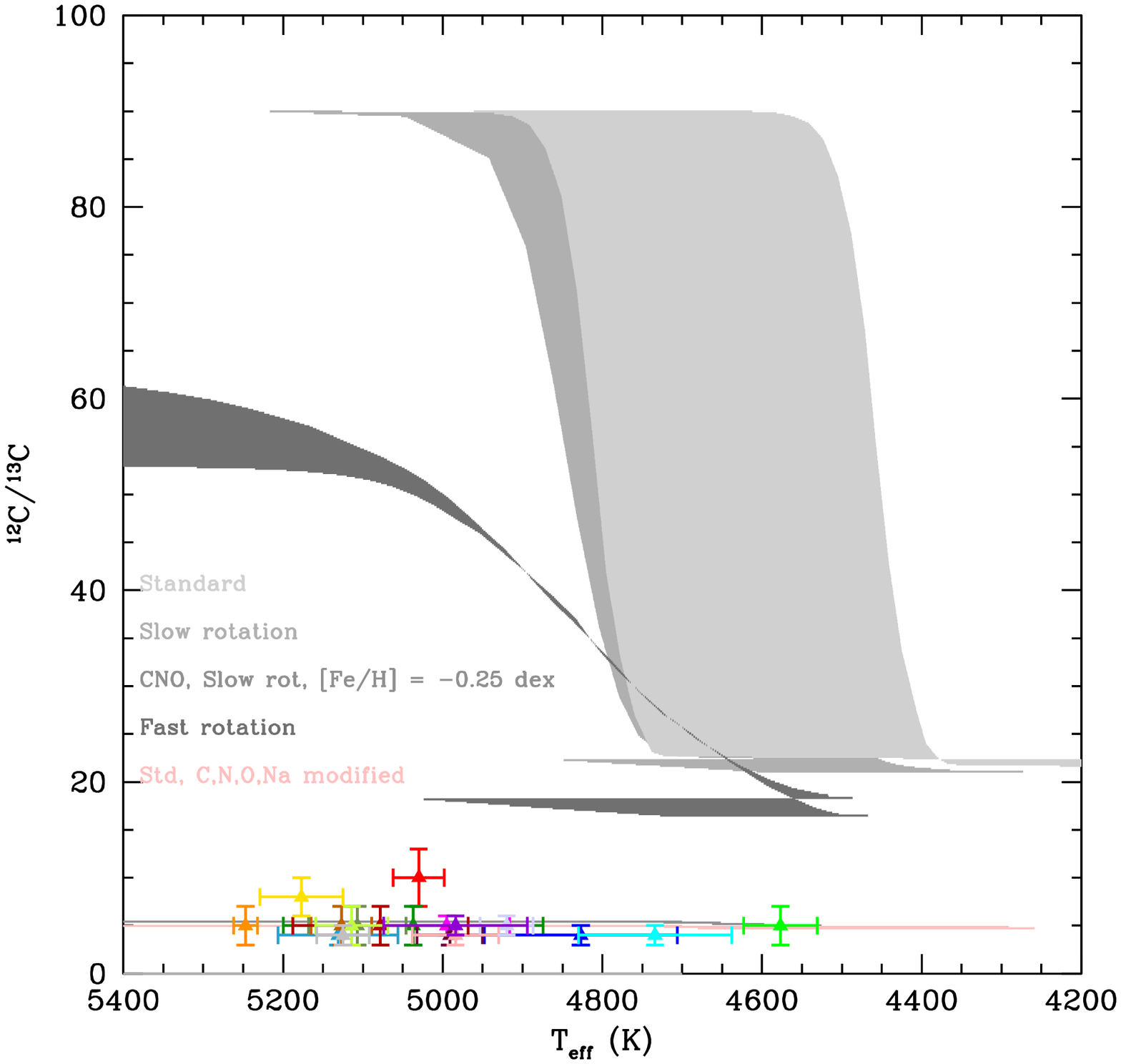}
\end{center}
%\vspace*{1.6cm}
\caption{A(Li) , [C/H], [N/H], and $^{12}C/^{13}C$ as a function of temperature in wGb stars compared to different model predictions.  The colours of the data points identify the stars as indicated in Table~\ref{tab_AP}. The shaded areas correspond to the regions of these planes that should be populated with points according to the predictions of the different families of models computed here as labelled in Table~\ref{tab_models}. In the upper left plot, the {\sf MildRot (Paper I)} label associated with the dark grey area indicates the predictions of rotating models with $\upsilon_{ZAMS} = 50$ \kms computed and presented in \citet{palaciosWGB2012}.}
\label{fig:abund}
\end{figure*}

 The carbon and nitrogen abundance patterns observed in wGb stars clearly indicate that the plasma is fully processed by the CN cycle. In stars in the mass range 3.2 to 4.2 M$_\odot$, hydrogen burning essentially occurs via the CNO cycle in the core during the main sequence evolution and in the hydrogen burning shell (HBS) in the subsequent evolution. According to standard stellar evolution theory, the abundances of $^{12}$C and $^{13}$C are expected to vary at the stellar surface during the 1st DUP, with a total decrease in [C/H] of 0.2 dex and in $^{12}$C/$^{13}$C from 90 to 20. Recently, \citet{AL14} have derived the carbon and oxygen abundances of a new sample of A-type stars across the Hertzsprung gap and confirm that no abundance anomaly develops during the evolutionary phase that connects the turn-off in the blue to the red giant branch in the red. Our non-rotating models are also consistent with this picture (see Fig.~\ref{fig:abund}). For slowly rotating stars, the surface carbon abundances are expected to be slightly modified at the end of the main sequence, but the overall impact of rotation induced mixing in stars with $\upsilon_{\rm ZAMS} \approx 25$ km.s$^{-1}$ is negligible as can be seen in Fig.~\ref{fig:abund} (the difference between the areas in the two lighter shades of grey). For rapid rotators on the ZAMS , despite the larger turbulent shear mixing triggered by the larger angular velocity, the transport is not efficient enough  to significantly decrease the surface carbon abundance prior to the 1st DUP as was shown in \citet{palaciosWGB2012}, and as can be seen from the black shaded areas in Fig.~\ref{fig:abund}.\\ Our rotating models include the same physcial processes and the same treatment as those proposed by \citet{CL10} to explain the Li and C underabundances and the N overabundances observed in lower-mass RGB stars of the Galactic field and of globular clusters. In solar metallicity intermediate-mass stars, the rotational mixing is not very efficient and the lowest predicted values of [C/H] $\approx -0.3$ dex and $^{12}C/^{13}C \approx 20$ for stars past the 1st DUP are far from the mean values <[C/Fe]> $\approx -1.2$ dex and <$^{12}C/^{13}C> \approx 4$ encountered in the wGb stars.
 
%% Figure 11
%% Figure 11

\subsubsection*{Nitrogen}

According to standard stellar evolution theory, the abundances of $^{14}$N is expected to vary at the stellar surface during the 1st DUP with a global increase of [N/H] by 0.5 dex. This prediction is met by our models and were not modified if wGb stars were slow rotators at the ZAMS as shown in the lower left panel of Fig.~\ref{fig:abund}. For rapid rotators on the ZAMS, the larger turbulent shear mixing allows a higher enrichment of the surface with nitrogen and a post-dredge-up value of about 0.6 dex (see the dark grey shaded area in Fig.~\ref{fig:abund}).\\ As in the case of carbon, the maximum overabundance reached in our models is much smaller than the mean value of <[N/Fe]> $\approx +1.3$ dex derived for the wGb stars.\\

\begin{figure*}[t]
\begin{center}
%\vspace*{-2cm}
\includegraphics[width=0.45\textwidth]{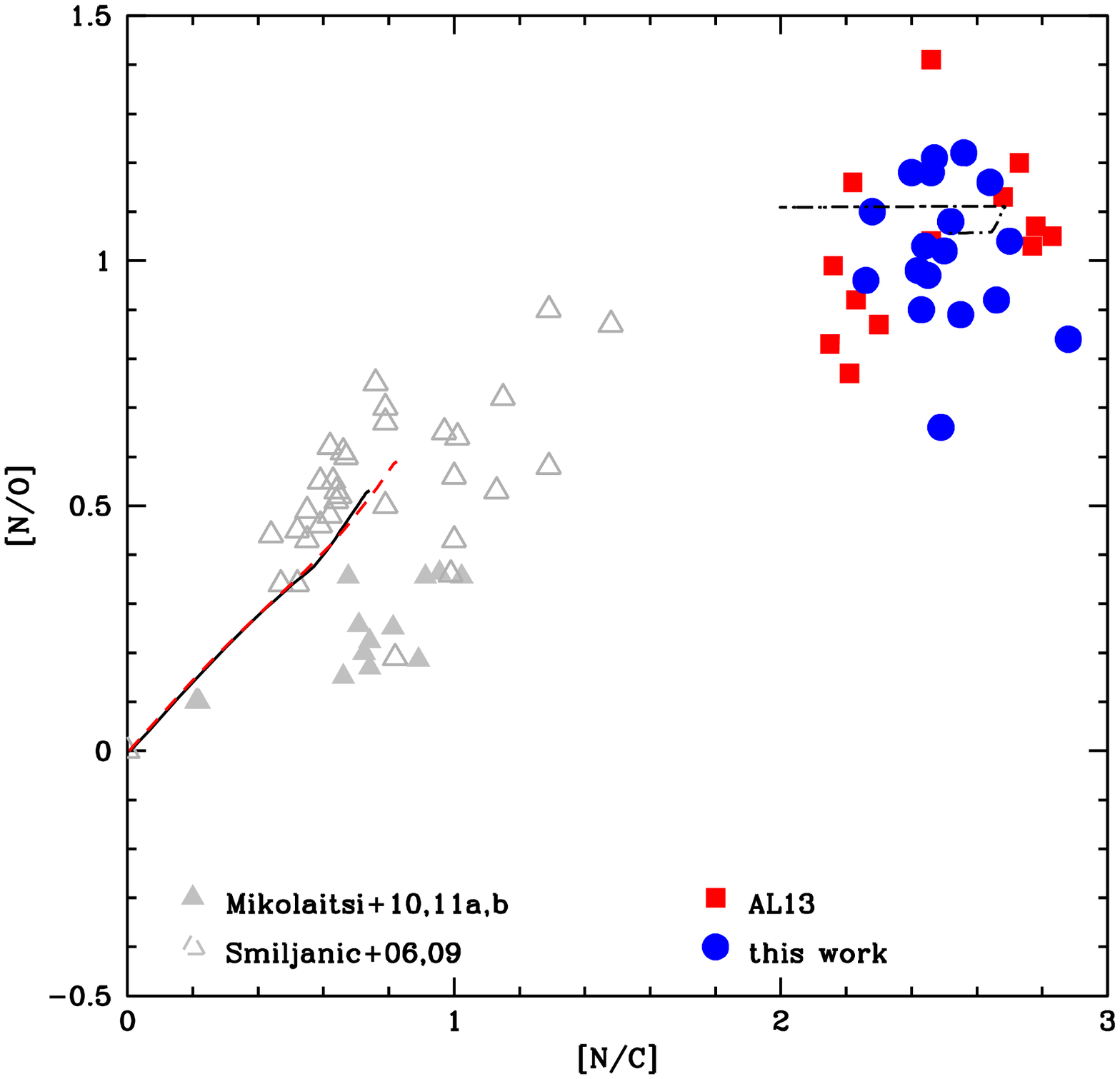}%
\includegraphics[width=0.45\textwidth]{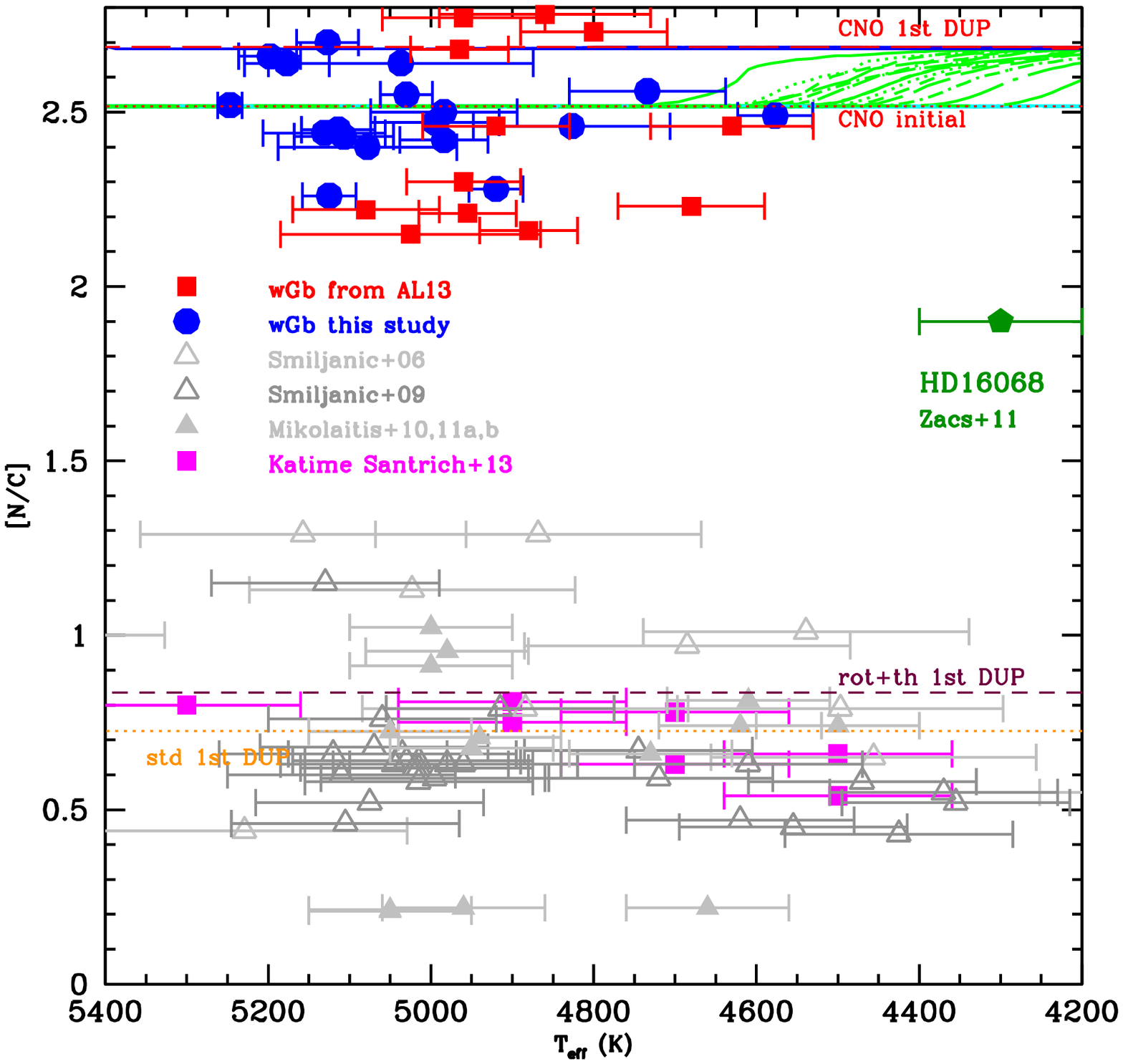}
\end{center}
%\vspace*{1.6cm}
\caption{ C, N, and O in the wGb stars and other intermediate-mass giants. \underline{\em Left}  [N/O] as a function of [N/C] for the stars in our sample (blue dots) that are in common with {\sf AL13} (red squares) compared with abundances in intermediate-mass stars of the field from \citet{Smiljanic06}, and of open clusters from \citet{Smiljanic09} (open grey triangles), \citet{Mikolaitis10,Mikolaitis11a,Mikolaitis11b} (filled grey triangles), and  \citet{Santrich13} (magenta filled squares). Also shown are the predictions from standard stellar evolution models of 3 M$_\odot$ and 4 M$_\odot$ including rotational and thermohaline mixing from \citet{Lagarde12} (solid black line and red dashed line in the lower left part of the plot) and from our {\sf CNO} models (short-long-dashed line). \underline{\em Right}  [N/C] as a function of temperature for the wGb stars and the same intermediate-mass red giants as on the left plot. An additional point, the giant HD 16068 from the solar metallicity open cluster Tr2, is also plotted after \citet{Zacs11}. The lines represent the expected post 1st DUP value of [N/C] predicted by our different models as labelled. For the {\sf CNO} models we also show the main sequence value of this ratio. }
\label{fig:compotherobs}
\end{figure*}

Fig.~\ref{fig:compotherobs} compares the CNO abundances in wGb stars and in other intermediate-mass giants of similar metallicities in the form of the [N/C] ratio. The very large anti-correlated  nitrogen overabundances and carbon underabundances in the wGb giants places them in exclusive regions of the [N/O] vs. [N/C] and [N/C] vs. \teff planes compared to other giants in the field and in young open clusters, sharing the same spectral type and luminosity class. Unlike other intermediate-mass stars found in young open clusters, models including rotational and thermohaline mixing cannot account for the observed abundances.\\ We note that the differences between the temperatures we derived and those derived  by {\sf AL13}  for the subsample of wGb stars common to both our studies do not alter the peculiarity of wGb stars, which  does not depend  on their evolutionary status as far as C and N abundances are concerned.

\subsubsection*{Heavier elements}

When occurring at high temperatures, the NeNa-chain is associated with the CNO cycle. This chain is made of a series of nuclear reactions that  lead to an overall increase in the number of $^{23}$Na nuclei at the expense of $^{22}$Ne. This increase occurs at constant number of $^{16}$O nuclei first, and then--when the temperature  increases further--it is accompanied by a decrease in the number of $^{16}$O nuclei.

%Figure 12
%Figure 12
 As the carbon and nitrogen abundance patterns observed at the surface of wGb stars indicate that their atmospheres have been processed through a hot CNO cycle,   sodium and maybe oxygen anomalies should also be expected at their surface. As shown by {\sf AL13} and noted in Table~\ref{tab_abund1}, wGb stars are indeed mildly enhanced with sodium ($<[Na/Fe]> \approx 0.35$ dex) and have oxygen abundances in good agreement with other field stars. \\ The heavier elements are normal compared to similar giants.
The abundances of O, Na, and heavier elements are thus fully consistent with those of C and N within the framework of the hot CN cycle, clearly indicating that the material exposed at the surface of the wGb stars has been nuclearly processed through the CNO cycle and the NeNa- chain.\\ We note that   at such temperatures Li is already not expected to survive, and so--as pointed out in \citet{palaciosWGB2012}--the presence of this light nuclide in the atmospheres of many of the wGb stars introduces a puzzling inconsistency in an otherwise very clear nuclear pattern.\\ In \S~\ref{sec:scenario}, we  examine the different ways such a nuclear pattern can be achieved in stellar atmospheres of evolved intermediate-mass stars.

\section{Exploring the different scenarios}\label{sec:scenario}

In the previous sections we  provide a factual report on the fundamental parameters; on the abundances of Li, C, N, O, Sr, and Ba;  and on the kinematics of the wGb stars in our sample. At first glance, our spectral and dynamical analysis indicates that 
\begin{enumerate}
\item wGb stars have an initial mass between 3.2 and 4.2~\msun. They are presently evolving either on the lower red giant branch or, less probably, on the red clump from progenitors on the main sequence that should have been late B- to early A-type stars;
\item the dynamical analysis of the wGb stars in our sample reveals no peculiar behaviour in terms of kinematics compared to the sample of normal giants by \citet{LH07};
\item almost 50\% of the wGb in our sample exhibit more lithium at their surface than expected from 1st DUP dilution predictions, indicating either that it has been only mildly destroyed or that they have undergone some sort of enrichment. Almost a third of the stars have very low lithium abundances, that can be interpreted as either post DUP abundances of fast rotating stars or as the signature of pollution with lithium poor material;
\item the carbon, nitrogen, oxygen, and sodium abundances analysis clearly shows the presence at the surface of the wGb stars of matter that was fully processed through the CN cycle and the NeNa- chain; %as also pointed out by {\sf AL13};
\item the abundances of s-elements Sr and Ba reveal no specific enrichment that could be associated with  pollution by more massive asymptotic giant branch (AGB) stars.
\end{enumerate}

Taking these points into account, we explore here the self- and external pollution scenarios that might be considered to account for the chemical peculiarities of this specific class of stars.

\subsection{Self-enrichment}

The self-enrichment scenario proposes that the abundance peculiarities observed at the surface of wGb stars result from internal mixing that brings to the surface the products of the CN cycle nucleosynthesis occurring in their deep interior. It is based on the fact that the abundances of the elements heavier than C at the surface of wGb stars indicate that the atmospheres of these stars have been processed by a hot CN cycle.
%Figure 13
\begin{figure*}[t]
\begin{center}
%\vspace{-1cm}
\includegraphics[width=0.3\textwidth]{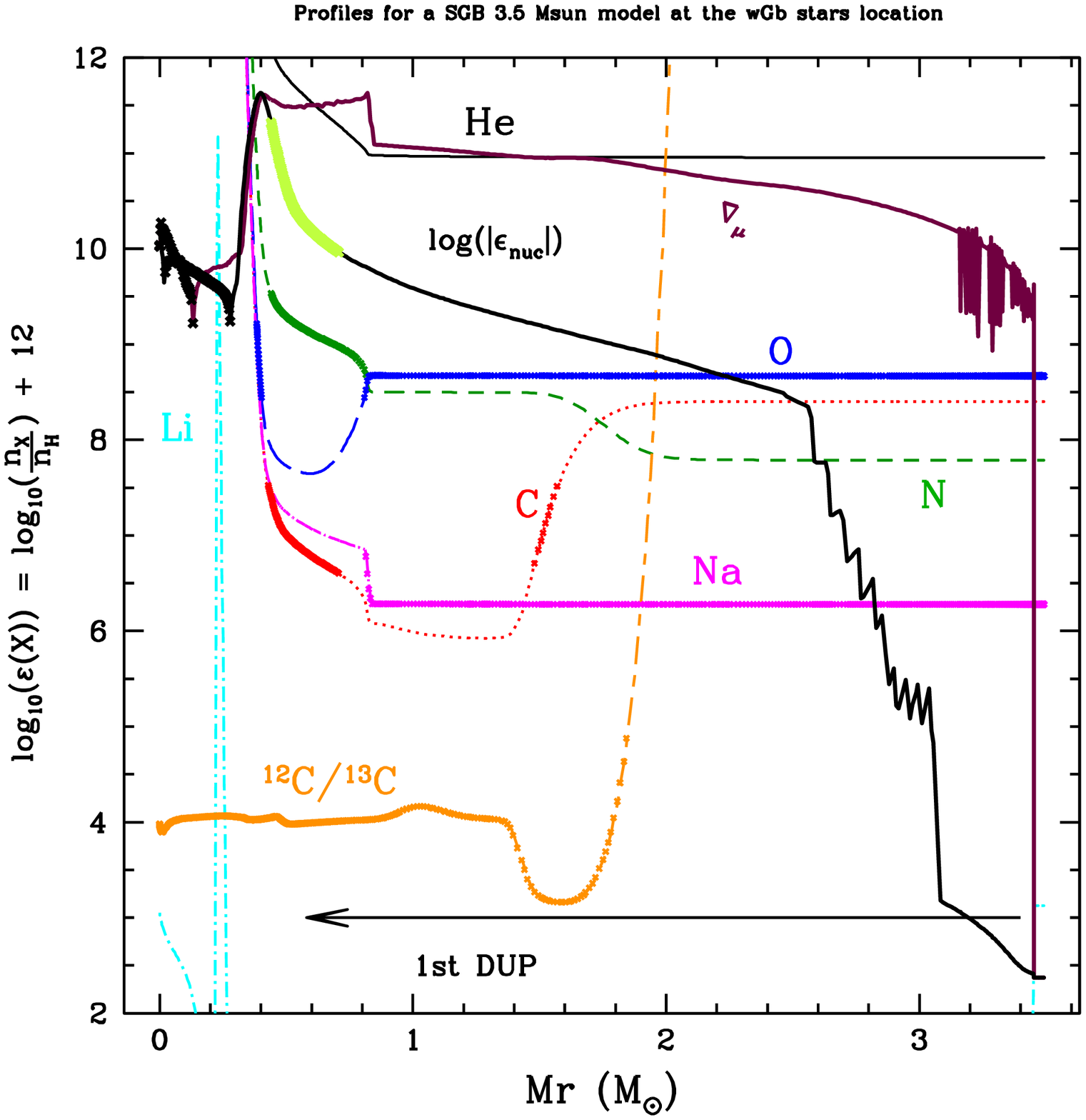}%
\includegraphics[width=0.3\textwidth]{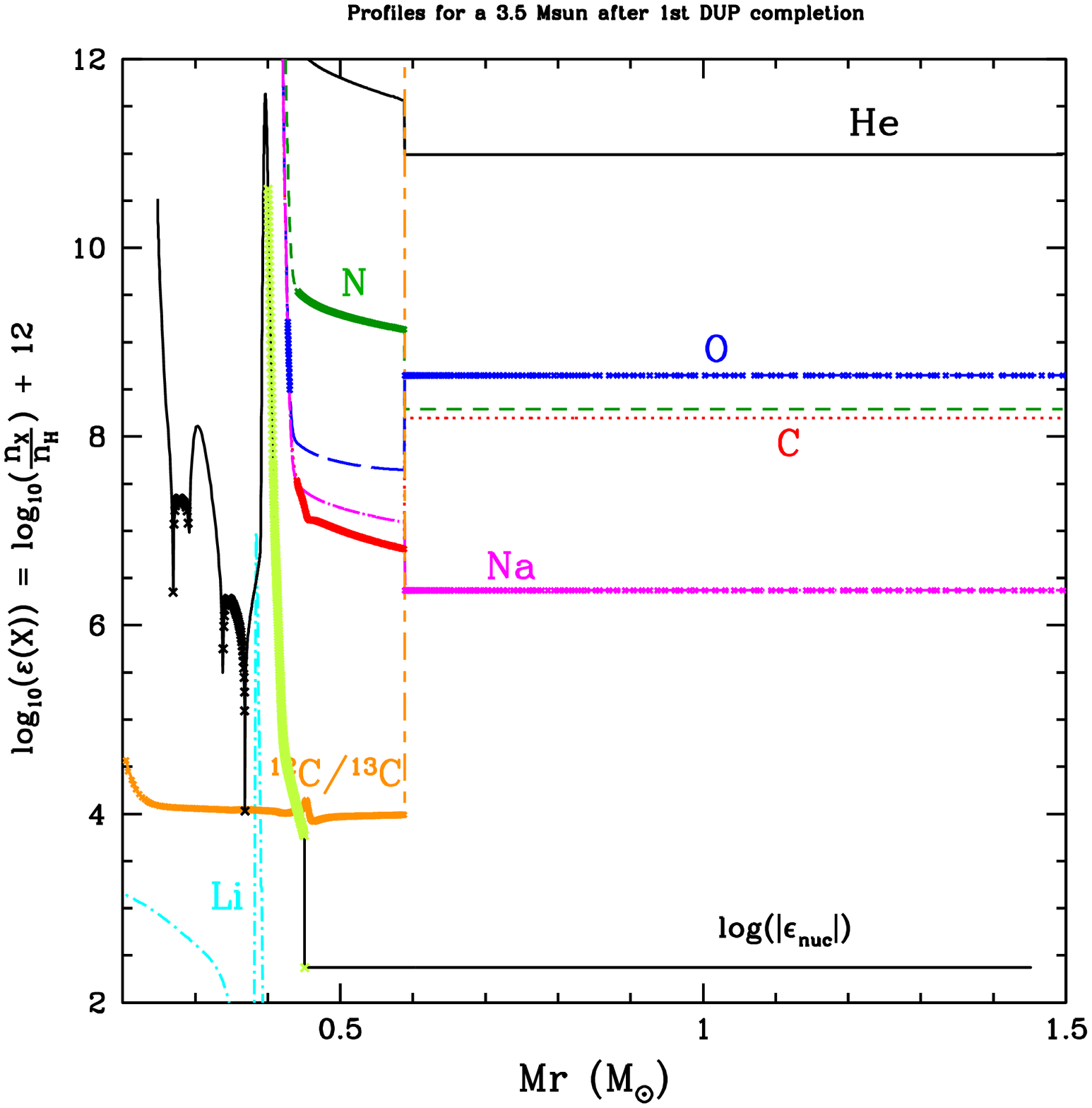}%
\includegraphics[width=0.3\textwidth]{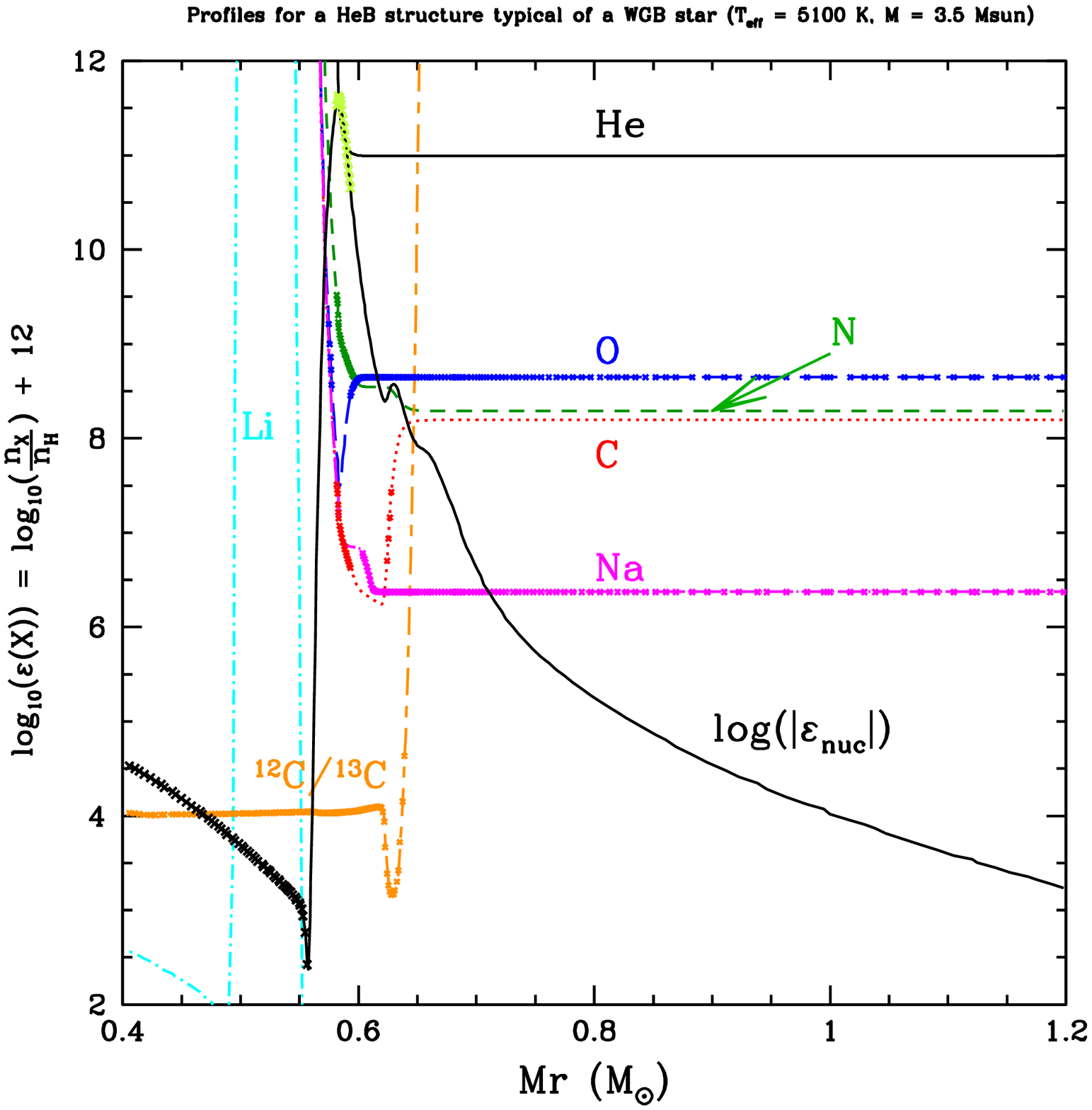}
\end{center}
%\vspace*{1cm}
\caption{Abundance profiles in the standard 3.5 \msun model at solar  on the SGB at the location of wGb stars ({\em
left}),  at the completion of the 1st DUP ({\em center}), and  in the clump at the location of wGb stars ({\em right}). The bold parts of the tracks correspond to the mean values derived for our sample of wGb stars. The bold black line represents the nuclear energy generation inside the star (the scale of which is not presented on the plot), and the lime-green bold part on this track represents the region where the surface C and N abundances of the wGb stars are encountered inside the model. }
\label{fig:Profiles}
\end{figure*}
%Figure 13
The CN cycle is part of the CNO bi-cycle and its main characteristic is the anti-correlation of carbon and nitrogen abundances with low carbon abundance and high nitrogen. 
When occurring at equilibrium, the carbon isotopic
ratio $^{12}$C/$^{13}$C $\approx$ 4. This nuclear cycle occurs inside
intermediate-mass stars in the mass range of the wGb stars, which lead
{\sf AL13} to propose self-enrichment to explain the C and N abundances at their surface. In Fig.~\ref{fig:Profiles}, we show the internal profiles of $^4$He, $^7$Li, $^{12}$C, $^{12}$C/$^{13}$C, $^{14}$N, $^{16}$O, and $^{23}$Na as a function of the mass coordinate in our standard 3.5 \msun model during the subgiant phase (left and middle panels) and the core helium burning phase (right panel). The bold parts of the tracks represent the domain of surface abundances of these nuclides derived for the wGb stars in our sample (and listed in bracket notation in Table~\ref{tab_abund1}). 
The nuclear energy profile is also plotted as a bold black line on each panel. In this figure, the left and the right panels correspond to the internal structure of a model with the luminosity and effective temperature within the domain covered by wGb stars, while the central panel shows the profiles at the completion of the 1st DUP.

The first thing to note is that the CNO cycle is actually operating with the expected amplitude inside the wGb stars.% as noted by {\sf AL13}. 
The second point is that the carbon and nitrogen surface abundances vary due to the 1st DUP, as shown in the middle panel presenting the profiles at the DUP completion. The flat parts of the profiles beyond $\approx$ 0.6 \msun are within the convective envelope. The corresponding carbon and nitrogen abundances result from the deepening of the convective envelope in regions depleted (resp. enriched) in carbon (resp. nitrogen) shown on the left panel of the same figure. The surface abundances of N and C do not vary during the He burning phase (see right panel).

As can be seen in this figure, the region where the CNO cycle operates at
full length has a small extent (in mass and radius) on the SGB and is
even shallower if the star is at the clump. 
This region is deeply buried in the core and it is associated with the
region where the peak of nuclear energy is produced. It is also fully depleted of lithium.

Assuming a self-enrichment scenario with an undefined cool bottom mixing process that would connect the convective envelope (its base  located around the mass coordinate $M_r \approx 3.45 \msun$ at the evolutionary stage shown in the left panel) to the region where the CNO cycle operates (at $M_r \approx 0.7 \msun$) implies that there would be a physical mechanism able to overcome the large mean molecular weight barrier (shown as a merlot reddish brown bold line in the left panel of Fig.~\ref{fig:Profiles}) associated with the HBS to reach  the shells where the CNO cycle operates at full length, which is where the bulk of the nuclear energy sustaining the star is produced.  Considering the distance between the base of the
envelope and the hydrogen burning shell ($\gg$ 15 \rsun), the
mixing process should be very efficient, equivalent to a uniform
diffusion coefficient of at least $10^{16}$ cm$^2$.s$^{-1}$, which is
larger than the mixing efficiency in convective regions. Such a transport process would also fully deplete the lithium in the envelope of the stars. \\The rotational transport included in stellar evolution codes (meridional circulation + shear turbulence + thermohaline) is strongly limited by the mean molecular weight barriers and cannot reach the demanded efficiency.

In order to estimate the expected impact of the self-polluting process, we ran some test computations (not shown here) adding an ad hoc diffusion coefficient mimicking the specificities of the previously
described cool bottom processing. The code becomes unstable and the
result is a strong deviation from the evolutionary path owing to the
mixing of the energy production region that affects the CNO cycle.

 We conclude that the requirements of a very efficient and deep cool bottom processing that would process the surface shells through the CNO cycle while preserving a non-zero lithium abundance and a normal evolution of the star in the HR diagram cannot be met considering the present knowledge of mixing processes in stellar interiors. This conclusion is independent of the assumed evolutionary status of the wGb stars.

\begin{figure}
\begin{center}
%\vspace{-2cm}
\includegraphics[width=0.45\textwidth]{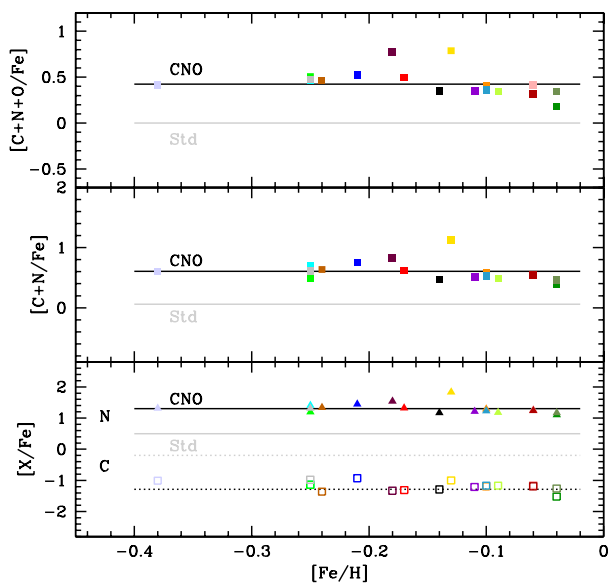}
\end{center}
%\vspace{1.4cm}
\caption{Abundances of C, N, C+N, and C+N+O derived for the wGb stars in our sample as a function of metallicity. The predictions of the post 1st DUP values (i.e. at maximum variation) from our standard solar chemical composition models and from our {\sf CNO} solar metallicity models are also shown as grey and black lines, respectively.}
\label{fig:CNOvsFe}
\end{figure}

 Finally, in our models (rotating and non-rotating), the sum [(C+N+O)/H] is conserved along evolution and is equal to 0 (Fig.~\ref{fig:CNOvsFe}), well below the average value of about +0.5 dex derived for the wGb stars in our sample. To reach such a large increase (more than a factor of 3) would require the production of large amounts of primary $^{14}$N deep in the star and the transport of this primary nitrogen in the atmospheres of wGb stars together with the H-burning ashes. If the production of primary $^{14}$N can be achieved in intermediate-mass stars during He-burning phase, it requires extreme conditions (very low metallicity, fast rotation, and advanced evolutionary status beyond the red clump) that are not met in wGb stars.\\
% Figure 14
% Figure 14

For these reasons, we conclude that the self-pollution scenario should not be considered as a valid option to account for the strong carbon deficiency (and nitrogen enrichment) observed at the surface of wGb stars. This conclusion is independent of any assumed evolutionary status for the wGb stars. \\
As already pointed out in \S~\ref{sec:abund} and in \citet{palaciosWGB2012}, the detection of lithium in the atmospheres of the wGb stars is  incompatible with the amount of mixing that would be required to account for the C, N, and Na abundance patterns. It is interesting to note that the Li-rich wGb stars that would be past the 1st DUP completion are as puzzling as the more normal, yet very rare lithium-rich red giants. To date, no consistent model is able to explain lithium  overabundances of red giants that have evolved beyond the 1st DUP \citep{Gonzalez2009,MS2013,Santi2015}.

\subsection{Pollution}

In the pollution scenario, the abundance pattern observed at the surface of the wGb stars results from the accretion of material processed in other objects (stars or planet-like objects) at an earlier epoch.

In close binary systems, the pollution can occur whenever mass overflows from a more evolved companion. In these cases the pollution concerns the outer layers only. If the primary is a low-mass main-sequence star with a convective envelope, the polluted material will be diluted in the surface convective region and should remain at its post-dilution level until a dredge-up event erases it by mixing the polluted layers with the internal pristine plasma. If the primary is an intermediate-mass or a massive star, then the outer layers are radiative and the polluted material that is accreted is not diluted. However, in these external layers, the nuclides are prone to atomic diffusion (gravitational settling and radiative levitation) and peculiar abundance patterns may develop. This is clearly shown by the wealth of chemically peculiar stars encountered among the A and B spectral type main-sequence objects.

Among the stars described in Table~\ref{tab_AP}, only HD 166208 (observed with NARVAL in the northern hemisphere) is identified as a spectroscopic binary. HD 49960, HD 67728, and HD 91805 are reported as part of double systems in the Catalog of Components of Double and Multiple Stars \citep{CCDM02}. The 14 other objects (73.6\% of our sample) should be single stars. Considering this information, we cannot attribute the carbon and nitrogen abundance anomalies to binarity in the case of the wGb stars.\\

Pollution may also occur in non-binary systems via planet accretion during the red giant phases. This particular scenario of pollution was proposed by \citet{SL99a,SL99b} as a possible explanation for the existence of Li-rich giants on the upper  RGB for solar-mass stars and the early-AGB for intermediate-mass stars.
\cite{Carlberg12} also studied the impact of planet accretion on the abundances and rotation of low-mass red giants and explain that planet accretion by red giant stars should lead to a noticeable increase in their rotation velocity due to the transfer of the planetary orbital momentum to that of the stellar envelope, and more importantly, to a light element replenishment, e.g. an increase in the surface lithium abundance and in the carbon isotopic ratio, and to an increase in the abundance of refractory elements. \\ The planet engulfing scenario is particularly appealing to account for high Li abundances in the Li-rich wGb stars, but it cannot account for the abundances of heavier nuclides (C and N anomalies and normal abundances from refractory nuclides as shown by {\sf AL13}).

Another possibility for the pollution scenario is the ab initio pollution with material that has been processed through the CNO cycle. We have explored this scenario with the computation of the models labelled {\sf CNO} and {\sf RotCNO} in Table~\ref{tab_models}, assuming that the wGb stars were either formed or polluted while still fully convective on the pre-main sequence by material bearing the imprint of CNO cycle processing. These models are computed assuming initial abundances of C, N, O, and Na as indicated in Table~\ref{tab_models}. In this approach we do not specify the origin of the pollution and we keep a normal initial lithium abundance of 3.2 dex, not taking into account the fact that from a nucleosynthetic point of view, CNO cycle processed material should be devoid of lithium. As can be seen in Figures~\ref{fig:HRLiSGB} and \ref{fig:HRLiclump}, these models evolve as standard models that would have a higher metallicity. Although the iron content is maintained solar, the combination of the initial abundances chosen for C, N, O, and Na 
 modify the total metal content and the heavy elements mass fraction $Z$ is increased by a factor of 2 compared to the standard models (see Table~\ref{tab_models}). The major features of these models is thus to appear as lower mass models on the main sequence and Hertzsprung gap, and to evolve on short blue loops during the core helium burning phase. As discussed earlier, using these tracks to determine the masses of wGb stars leads to masses higher than those given in Table~\ref{tab_MRage} by $\approx$ 0.2 \msun. Concerning the evolution of lithium, their surface abundance is similar at all points to the standard value (see Figs.~\ref{fig:HRLiSGB} and \ref{fig:HRLiclump}), so that similar models including rotation will lead to a good agreement between a location on the Hertzsprung gap, the lithium content, and the carbon and nitrogen abundances. As shown in Fig.~\ref{fig:abund}, these models--designed to reproduce the carbon and nitrogen abundances--are the only ones able to reconcile all the observables.
As expected by construction and as shown in Fig.~\ref{fig:CNOvsFe}, the surface abundance variation of CNO elements following the 1st DUP obtained in these models is in very good agreement with the values derived in the wGb stars in our sample.\\

\subsection{Caveats of the pollution scenario}

The pollution scenario as illustrated by the  {\sf CNO} and {\sf RotCNO} models, faces the same weakness as the self-enrichment scenario when it comes to reconciling the surface abundances of Li-rich wGb stars and those of the CNO nuclides. In both cases the CNO pattern of wGb band stars is the result of CNO cycle nucleosynthesis occurring  at high temperature where lithium is fully destroyed by proton captures in stellar interiors.\\

Assuming that the C and N abundance anomalies existing in wGb stars result from an initial / early pollution, two other major questions appear: (1) Can we find earlier-type progenitors and successors of wGb stars? (2) What type of stars can produce the pollution we have identified ?\\

(1) Considering the masses of the wGb stars, they were early A-type to
late B-type stars on the main sequence. Because such stars are
devoided of a convective envelope and do not experience the same high
rate of mass loss as the O-type stars many of them -- if not all-- are
classified as chemically peculiar stars due to the combined action of
hydrodynamical mixing processes and atomic diffusion (gravitational
settling, radiative levitation) in their atmospheres and upper
envelopes. The identification of potential progenitors to wGb stars is
thus very difficult, and no clear evidence of dwarf stars exhibiting
similar CNO abundance patterns to the wGb giants has been
found.\\ Turning to the possible descendants of wGb stars, considering
that a large part of the sample are evolving in the Hertzsprung gap,
the wGb stars that are Li-poor and compatible with a clump
evolutionary status would be their descendants. Indeed, the surface CNO abundance of wGb stars in the pollution scenario are already close to nuclear equilibrium on the main sequence, and will be very marginally affected by the 1st DUP. Therefore, we expect the descendants of wGb stars in the pollution scenario to share similar CNO abundances with their progenitors, but to exhibit lower lithium surface abundances as a consequence of the 1st DUP. \\

(2) A review of the latest yields from stars more massive than 4 ~\msun indicates that none is expected to eject matter processed through the CN cycle exclusively at the end of its evolution \citep{WW95,FC97,Siess2010,CL13,Pignatari13,KL14}. The {\em massive} intermediate-mass stars and the massive stars evolve through He burning and beyond, and their final contribution to the enrichment of the interstellar medium bears the imprint of these nucleosynthetic processes that tend to erase the direct signature of H burning via the CNO cycle.\\ Another possibility could be that the wGb stars accreted matter from the winds of main sequence massive stars while they were still fully convective on the pre-main sequence. Such scenarios have been proposed at low metallicity to account for the abundance anomalies in globular cluster RGB stars \citep{PC06}. In the specific context of globular clusters, the slow equatorial winds of extremely fast-rotating low-metallicity massive stars is invoked, but there is no clear indication to date  of the existence of such objects at solar metallicity.  \citet{Martins15} recently published a study of ON stars, which are main sequence O-type stars presenting a strong N enrichment, but they still fall short in terms of the amplitude of the anomalies, getting [N/C] $\approx 2$ dex and [N/O] $\approx 0.5$ dex, while the mean for the wGb stars is [N/C] $\approx 2.5$ dex and [N/O] $\approx 1$ dex (see Fig.~\ref{fig:compotherobs}).

\section{Conclusions}\label{sec:conclusion}

Using high-resolution and high S/N spectra, we have analysed a sample of 19 wGb stars for which we derived the fundamental parameters, the metallicity, and the surface abundances of Li, C, N, O, and some heavy elements. Part of our sample was also analysed in {\sf AL13}, and despite some systematic differences between the fundamental parameters derived in both studies that are discussed in the light of the differences in the spectral synthesis and model atmospheres used, we confirm the results of {\sf AL13} concerning the strong deficiency in carbon, a carbon isotopic ratio at nuclear equilibrium, and a strong enrichment in nitrogen for all the wGb stars in the sample. This proves the presence of CN-cycle processed material in their atmospheres as also pointed out by {\sf AL13}. Based on a series of dedicated stellar evolution models, we find that wGb stars are most likely the descendants of early A- to late B-type stars, with initial masses between 3 and 4.2 \msun. A large fraction of them could be presently undergoing the 1st DUP in the red part of the Hertzsprung gap; some stars in our sample are more compatible with a clump status. While the lithium abundances found cannot be qualified as anomalous since they reflect the normal evolution of rotating subgiant stars prone to rotating mixing, the carbon and nitrogen abundances remain puzzling. The carbon and nitrogen abundances are  found to be fully compatible with the expected CNO cycle nucleosynthesis that also occurs within the wGb stars. However,  we discard an internal origin for these abundance anomalies based on the argument that the nucleosynthesis region of interest is located very deep within the HBS, shielded by a high mean molecular weight gradient, and corresponds to the region where the peak of nuclear energy is produced. Any mixing in this region would result in an abnormal evolution and evolutionary path of the stars in the HR diagram, which is not supported by observations.\\ We have reviewed the possibilities of an external pollution, and are not able at present to put up a consistent scenario that would explain the overall abundance pattern of wGb stars. They could have been born polluted or have been polluted in their infancy (while fully convective) by more massive stars that would have ejected CNO cycle processed material. However,  at the time of writing we cannot clearly  identify  the polluters nor fully reconcile the normal lithium evolution with a pollution with CN-cycle processed material that should be, by nature, devoid of lithium. Obtaining new measurements of Be for the wGb stars would add a valuable constraint and clarify the Li abundances with respect to the CNO.\\
Considering that little is known about the binary status of wGb stars in our sample, we have started a follow-up campaign on selected apparently single objects with the HERMES spectrograph on the Mercator Telescope in order to clearly establish their nature.\\ We suggest that further studies of more massive stars should be done in order to identify the possible polluters of wGb stars. A careful analysis of large surveys in search of new wGb stars would also be of great help.

\begin{acknowledgements}
The authors thank the anonymous referee for help in improving the overall quality of the paper. AP would like to thank O. Richard, F. Martins, and C. Charbonnel for fruitful and interesting discussions. We are also thankful to Caroline Soubiran and R\'emi Cabanac 
for the observations of two stars at the Pic du Midi with the TBL telescope. Part of the work was done during  MP's visiting professor position in IUCAA.
MP is thankful to the director of IUCAA for his kind support. This research has made use of the SIMBAD database, operated at
CDS (Strasbourg, France) and of NASA's Astrophysics Data System Service.
\end{acknowledgements}

\bibliography{WGBpaper}

\appendix
\section{Observations log}
\begin{sidewaystable*}
\caption{Observation log of the target stars observed in this paper. Coordinates, V magnitude, and spectral types are from SIMBAD. TBL corresponds to the 
2.0m  Bernard Lyot Telescope installed at the Pic du Midi (France), and ESO corresponds to the ESO 2.2m telescope installed at La Silla (Chile). 
Accuracy on barycentric radial velocities (RV) is better than 0.07km.s$^{-1}$ for all stars.}
\centering
\begin{tabular}{ccccccccccc}
\hline
Identifier & RA & DEC & V & Spectral Type & Observatory & Date  & TU & Exposure Time & SNR  &  RV \\ 
&  (2000) &  (2000) & &&&YYYY-MM-DD& HH:MM:SS & (s) && (km.s$^{-1}$) \\
\hline
\hline
HD 18474 &  02 59 49.79190 & +47 13 14.4814 & 5.50 & G5:III...       & TBL  & 2011-10-03 & 02:01:45 &300   &   400 &     0.78  \\
HD 49960 &  06 49 40.54488 & -31 15 32.8492 & 8.35 & K2/K3IIIp+...   & ESO  & 2012-05-12 & 22:47:41 &1800  &   260 &    55.07  \\
HD 56438 &  07 14 36.40889 & -47 08 05.8079 & 8.11 & K1IIICNp...     & ESO  & 2012-05-12 & 23:50:16 &1800  &   300 &   -11.81  \\
HD 67728 &  08 08 37.64968 & -19 50 23.6865 & 7.54 & G9III           & ESO  & 2012-05-12 & 23:21:46 &1500  &   360 &     9.19  \\
HD 78146 &  09 05 44.21193 & -28 04 59.4164 & 8.57 & K1IIICNpv...    & ESO  & 2012-05-14 & 00:33:56 &2000  &   250 &    13.32  \\
HD 82595 &  09 31 57.7825  & -36 51 34.317  & 8.20 & K2II/III        & ESO  & 2012-05-13 & 01:27:59 &2400  &   260 &     8.78  \\
HD 91805 &  10 35 10.41242 & -43 39 52.5170 & 6.11 & G8II-IIICNpv    & ESO  & 2012-05-14 & 03:42:23 &300   &   320 &     8.89  \\
HD 94956 &  10 57 23.99200 & -29 16 50.7187 & 8.46 & K0IIICNpv       & ESO  & 2012-05-12 & 01:58:31 &3600  &   420 &   -77.43  \\
HD 102851 & 11 50 16.39738 & -51 28 14.3231 & 8.79 & K0/K1IIICNp...  & ESO  & 2012-05-14 & 02:40:43 &1500  &   340 &   -59.36  \\
HD 119256 & 13 43 36.28093 & -57 35 23.9100 & 7.33 & K1IICNp...      & ESO  & 2012-05-13 & 03:04:14 &1200  &   270 &     6.02  \\
HD 120170 & 13 47 58.7904  & -08 47 22.733  & 9.03 & K0              & ESO  & 2012-05-14 & 06:19:41 &305   &   170 &    12.50  \\
HD 120213 & 13 55 38.88086 & -82 39 58.2768 & 5.96 & K2.5III:p       & ESO  & 2012-05-14 & 03:33:06 &300   &   200 &   -32.97  \\
HD 124721 & 14 16 24.5544  & -45 11 23.125  & 9.48 & K1/K2III:       & ESO  & 2012-05-12 & 05:52:56 &2400  &   240 &   -40.45  \\
HD 146116 & 16 14 38.39564 & -00 23 55.4940 & 7.70 & G7III           & ESO  & 2012-05-12 & 06:34:36 &1800  &   330 &    31.83  \\
HD 165462 & 18 06 07.42766 & -00 26 46.0587 & 6.35 & G8IIp           & ESO  & 2012-05-14 & 05:26:46 &300   &   340 &    -9.19  \\
HD 165634 & 18 08 04.97982 & -28 27 25.5316 & 4.57 & G7:IIIb         & ESO  & 2012-05-14 & 03:51:11 &100   &   100 &    -4.12  \\
HD 166208 & 18 07 28.73963 & +43 27 42.8438 & 5.01 & G8IIICN...      & TBL  & 2012-09-08 & 21:02:16 &120   &   130 &   -16.63  \\
HD 204046 & 21 26 54.12950 & -33 54 35.6809 & 9.04 & G9IIICNpv       & ESO  & 2012-05-14 & 07:44:55 &2400  &   320 &    22.02  \\
HD 207774 & 21 51 33.58431 & -07 24 14.2895 & 8.89 & G5              & ESO  & 2012-05-12 & 09:14:22 &2400  &   280 &    11.92  \\
\hline
\end{tabular} 
\label{tab_log_observations}        
\end{sidewaystable*}

\end{document}